\documentclass[journal]{IEEEtran}

\usepackage[pdftex]{graphicx}
\usepackage{amsmath,bm}
\usepackage{amssymb}
\usepackage{color,soul}

\usepackage{amssymb}

\usepackage{multicol}
\usepackage{booktabs}
\newtheorem{Remark}{Remark}

\usepackage{cases}
\usepackage{multirow}
\usepackage{tablefootnote}
\usepackage{cuted}

\usepackage[noadjust]{cite}

\graphicspath{{figures/}}

\begin{document}
\title{Low-Power Control of Resistance Switching Transitions in First-Order Memristors}

\author{Valeriy~A.~Slipko, Alon~Ascoli,~\IEEEmembership{Senior~Member,~IEEE}, Fernando~Corinto,~\IEEEmembership{Senior~Member,~IEEE}, and Yuriy~V.~Pershin,~\IEEEmembership{Senior~Member,~IEEE}% <-this % stops a space
\thanks{V.~A.~Slipko is with the Institute of Physics, Opole University, Opole 45-052, Poland (e-mail: \mbox{vslipko@uni.opole.pl})}
\thanks{A.~Ascoli and F. Corinto are with Department of Electronics and Telecommunications
Politecnico di Torino, Turin, Italy}
\thanks{Y.~V.~Pershin is with the Department of Physics and Astronomy, University of South Carolina, Columbia, SC 29208 USA (e-mail: \mbox{pershin@physics.sc.edu})}
\thanks{Manuscript received August ..., 2025; revised  ....}}

\maketitle

\begin{abstract}
%In many cases, the behavior of physical memristive devices can be relatively well captured by using a single internal state variable.
This study investigates the low-power control of resistance switching transitions in
%first-order
memristive devices described through a single state variable. A unique yet general approach, enabling to derive the most energy-efficient protocols for programming their resistances, is proposed. This low-power control paradigm is applied to a couple of differential algebraic equation sets, capturing the nonlinear dynamics of voltage-controlled devices.
Depending upon intrinsic physical properties of a memristive device, captured in the model formulas and parameter setting, and upon constraints on programming time and operating voltages, the optimal protocol may require the application of either a single square voltage pulse of height set to a certain level within the admissible range across a certain fraction of the programming time, or some more involved voltage stimulus of unique polarity, including trains of square voltage pulses of different heights, over the entire programming time.  The practical implications of these research findings are significant, as the development of energy-efficient protocols to program memristive devices
is a subject under intensive and extensive studies across the academic community and industry.
\end{abstract}

\begin{IEEEkeywords}
Memristors, Memristive Systems, Resistance Switching Transitions, Optimal Control, Functional Optimization
\end{IEEEkeywords}

\IEEEpeerreviewmaketitle

\section{Introduction} \label{sec:Intro}

Resistive Random Access Memory (ReRAM) cells %\cite{asc_adv_ele_mat_2022}
\cite{ielmini_waser_2016}
-- a broad class of memristive devices -- are currently attracting significant attention in the international academic community, as well as in the industry, because of their promising potential to form the next generation of information storage devices, which incorporate information processing capabilities across the very same physical media as well.
The spectrum of their electronics applications includes the design of non-volatile memories, the hardware realization of in-memory computing~\cite{borghetti2010memristive,8672912,pedretti2021memory} or mem-computing~\cite{di2013parallel,diventra2022,ascoli2020} paradigms,
and the development of bio-plausible neuromorphic systems~\cite{pershin2010experimental,moon2019rram,9618724}. %,ascoli2025}.
Like all resistive devices,  memristive devices dissipate energy. From a practical standpoint, \emph{reducing Joule losses while programming memristive devices is extremely important.} Recently, Pershin and Slipko introduced strategies for the low-power control of ideal and non-ideal memristive devices~\cite{slipko2024reduction} by employing standard techniques from optimal control theory~\cite{alekseev2013optimal}. Furthermore, the study presented in~\cite{astin2025low} explored optimal control strategies for fractional-order memristors. The methods proposed in~\cite{slipko2024reduction} are quite general, but their implementation, requiring the solution of the differential algebraic equations (DAEs), governing the nonlinear dynamics of memristors, can be challenging, especially in view of their typical complexity and analytical intractability.

From a practical viewpoint, in many cases, using a single internal state variable~\cite{chua76a} is
sufficient to capture the essential features of the response of a physical memristor device to input/initial condition combinations of interest (in view of the work reported in ~\cite{kim2020experimental}, the term memristor is used here to refer to a memristive device or system). %This is why first-order memristive models are prevalent in the literature.
The most general DAE set~\cite{chua76a}, defining a first-order voltage-controlled memristor, featuring an internal state variable, indicated with the symbol $x$, reads as
\begin{eqnarray}
I(t)&=&G_M\left(x,V \right)V(t), \label{eq:1}\\
\dot{x}&=&f\left(x,V\right), \label{eq:2}
\end{eqnarray}
where $V$ represents the voltage across the device, while $I$ denotes the current through it.
Eq.~(\ref{eq:1}) is a generalized state- and voltage-dependent form of Ohm's law, where $G_M\left( x, V\right)$ represents the %so-called
device \emph{memductance}, an abbreviation for its memory conductance. % $x$ is the internal state variable,
Eq.~(\ref{eq:2}) is an ordinary differential equation (ODE), referred to as \emph{state equation}, where $f(x,V)$ denotes the \emph{state evolution function}, dictating
the time evolution of the internal state variable under any input $V(t)$ of interest.
In the most common bipolar voltage-controlled memristors, stimuli of opposite polarities are employed to trigger RESET and SET transitions, respectively.

Keeping the focus on memristive devices, amenable to a first-order mathematical description, in this work we introduce a novel and effective change of variables in the Joule heat functional, which allows to reduce the solution of the constrained optimization problem to the determination of solutions for transcendental equations, which poses weaker challenges than the integration of DAE sets.
While the proposed theoretical methodology, based on Pontryagin's principle~\cite{alekseev2013optimal}, may be applied to the mathematical description of any voltage- or current-controlled memristor of order one,  in the main (appendix) we employ the voltage-controlled
voltage threshold adaptive memristor -- VTEAM --
model~\cite{Kvatinsky2013} (dynamic balance model~\cite{Miranda20a}) to cover a wide range of devices.

This paper is organized as follows.
% da qui
Sec.~\ref{sec:2} introduces the proposed memristor switching energy minimization approach, as well as the VTEAM non-volatile memristor model~\cite{Kvatinsky2013},
wherein the state evolution function depends on the voltage polynomially. The results of the application of the proposed optimization strategy to the VTEAM model
under unconstrained and constrained resistance switching scenarios are presented in Sec.~\ref{sec:3}, and further discussed in Sec.~\ref{sec:4}.
Conclusions are summarized in Sec.~\ref{sec:5}.
In the Appendix, the most energy-efficient resistance switching protocol is derived for a device
described by the dynamic balance model~\cite{Miranda20a},
wherein the state evolution function depends on the voltage exponentially.
While, rigorously speaking, this model falls in the class of volatile memristors, two of its parameters
are chosen so that it would approximate
the dynamics of a \emph{quasi-non-volatile memristor} (see the Appendix for more details).

\section{Methods} \label{sec:2}

\subsection{Control Strategy for Memristor Switching Energy Minimization} \label{sec:2a}

Our goal is to find the most energy-efficient switching
control protocol to minimize Joule losses for inducing a monotonic change in the internal state of a memristive device, defined through the system~(\ref{eq:1})-(\ref{eq:2}),  %switched
%changed monotonically
from an initial value, say $x_i$, to a \textcolor{black}{different} final value, say $x_f$.
\textcolor{black}{This rules out the choice of a permanently-null control voltage as the optimal solution for the optimization problem.}
Mathematically, the problem is stated as the minimization of a \emph{Joule heat functional}, i.e. via
%Joule heat functional
\begin{equation}
Q=\int\limits_{t_i}^{t_f}G_M(x,V(t))V^2(t)\textnormal{d}t \;\; \rightarrow \;\; \textnormal{min}, \label{eq:Joule}
\end{equation}

\noindent where $Q$ denotes the switching
%programming
energy, measured in Joule, $x$ is a function of time, corresponding to the solution of the state Eq.~(\ref{eq:2}), $x(t_i) = x_i$, $x(t_f) = x_f$, while $t_i$ and $t_f$ are the initial and final time instants of a pre-defined
%given
%\textcolor{black}{maximum allowable}
programming time interval $T$,
%(phase),
respectively.
Eq.~(\ref{eq:Joule}) may be further complemented by some constraints, if necessary, as discussed shortly.
Importantly, for a memristor model endowed with a single internal state variable,
%as considered in this work,
a proper transformation of variables can be applied to the Joule heat functional in Eq.~(\ref{eq:Joule}), which allows to simplify the determination of the solution to the minimization problem as compared to the approach discussed in~\cite{slipko2024reduction}.
Specifically, with the substitution
$\textnormal{d}t=\textnormal{d}x/f(x,V)$, which uses the state Eq.~(\ref{eq:2}), we transform the integrand in Eq.~(\ref{eq:Joule}) in such a way to turn the variable of integration from time to state. %$x$.
The new Joule heat functional under minimization boils down to

\begin{equation}
Q=\int\limits_{x_i}^{x_f}\frac{G_M(x,V(x))V^2(x)}{f(x,V(x))}\textnormal{d}x ,
%\;\; \rightarrow \;\; \textnormal{min},
\label{eq:Q2}
\end{equation}
where the voltage $V$ is now a function of $x$.
In addition, the time $t$ is expressed as a function of the state $x$ via
\begin{equation}
t(x)=t_i+\int\limits_{x_i}^{x}\frac{\textnormal{d}u}{f(u,V(u))}. \label{eq:5}
\end{equation}

\begin{figure}[tb]
\centering
\includegraphics[scale=.4]{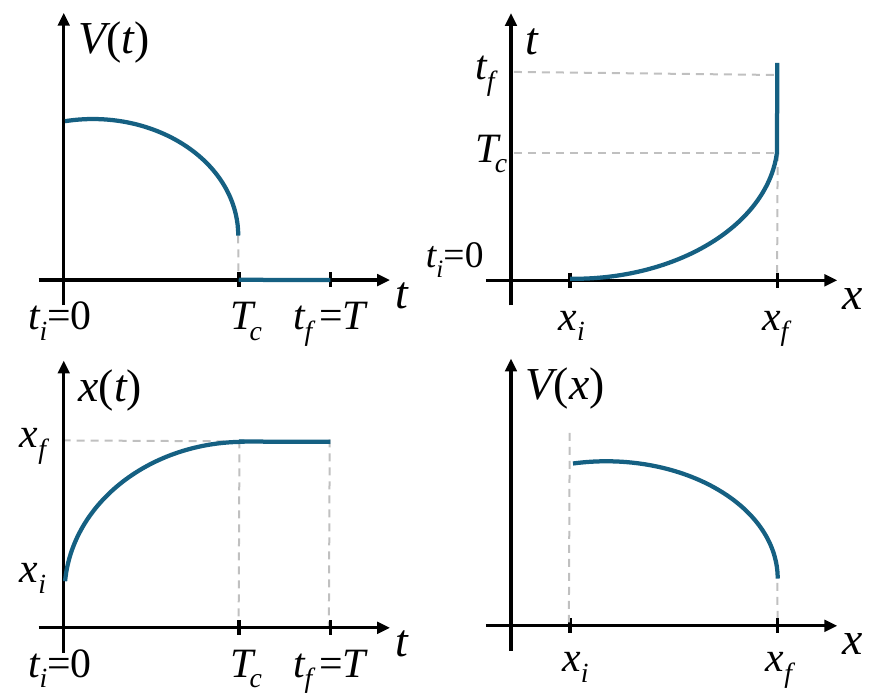}
\caption{Schematic illustration of a scenario, where $f(x,V)$ becomes equal to $0$ within the %$T$-long
programming phase, for an arbitrary non-volatile memristor model.
%\textcolor{red}{
%In these circumstances
%Here
The original time interval $[t_i,t_f]$, of duration equal to a pre-defined programming time $T=t_f-t_i$, is made up of a first time interval $[t_i,t_s]$, of width $T_c=t_s-t_i<T$, where %the state evolution function
$f(x,V)$ is non-zero,
featuring a unique sign -- assumed here to be positive (negative) for $V>(<)0$ so as to induce a monotonic increase (decrease) in the state $x$ --
%in the pedagogical example under consideration,
and the device %in principle
undergoes a resistance switching transition, \textcolor{black}{which may be of SET (RESET) or RESET (SET) nature, depending on the physical origin of its state $x$}, and of a second time interval $(t_s,t_f]$, of width $T-T_c$, where % the state evolution function
$f(x,V)$ is identically null, %equal to $0$,
%and %the memristor state
and $x$ keeps unchanged, as is the case for non-volatile memristors. As for all the simulation results, shown in this paper, we assumed $t_i=0$, resulting in the identity $T = t_f$.}
%but this is not necessary (see the Appendix).}
%For simplicity,
%As is the case for all the examples %presented
%in the paper, we assume $t_i=0$, resulting in the identity $T \equiv t_f$.}
%when $t_c<T$.
%Here, the switching interval $[0,t_c\}$ is followed by a zero voltage interval $\{t_c,T]$ in which the internal state is constant.
\label{fig:sketch1}
\end{figure}

\textcolor{black}{Importantly, Eq. \eqref{eq:Q2} requires the state evolution function $f(x,V)$ to be different from $0$ across the entire time interval, across which the memristor state $x$ is subject to a monotonic change from $x_i$ to $x_f$ under the application of a suitable non-zero voltage stimulus $V$ of fixed polarity across the device.}
\textcolor{black}{Let us allow the possibility for the state evolution function $f(x,V)$ to keep non-zero, maintaining a unique sign, only over an initial $T_c$-long temporal window $[t_i,t_s]$ of the pre-defined $T$-long programming time interval $[t_i,t_f]$, while vanishing thereafter. This eventuality may occur in non-volatile memristors.
Under these circumstances, a suitable voltage stimulus $V$ of unique polarity would program the memristor from an initial state $x_i$ to a different final state $x_f$ within a time frame shorter than the programming time $T$. \textcolor{black}{For this reason
%Here, in all respects,
$T_c$ %would %actually
%effectively
%represent the %actual
shall be referred to as the device \emph{effective switching time} in the remainder of the paper.}
In the example, referring to an arbitrary non-volatile memristor, illustrated in Fig. \ref{fig:sketch1}, a strictly-positive voltage signal $V$ induces a monotonic increase in the respective state from $x_i$ to $x_f$ within a time window of width $T_c$ shorter than a pre-defined programming time $T$.}

To allow scenarios of this kind, the following inequality-based constraint, descending from Eq. (\ref{eq:5}) must be included in the optimization process:

\begin{equation}
\int\limits_{x_i}^{x_f}\frac{\textnormal{d}x}{f(x,V(x))}\equiv
%\textcolor{red}{T_c}
\textcolor{black}{T_c =t_s-t_i \leq T =t_f-t_i,}
\label{eq:5a}
\end{equation}

\noindent where %, as anticipated above,
%$T \triangleq t_f-t_i$, and
the equality sign applies in the case, where
$t_f \equiv t_s$.

In integral form, the constraint (\ref{eq:5a}) may be re-written as

\begin{equation}
\beta[V(x)]=\int\limits_{x_i}^{x_f}\left(\frac{1}{f(x,V(x))} -\frac{T}{x_f-x_i}\right)\textnormal{d}x \leq 0, \label{eq:5b}
\end{equation}

\noindent which defines a new functional referred to as $\beta[V(x)]$.

For positive (negative) voltage-induced transitions, we assume that the non-zero control voltage lies in the interval
$V_1 \leq V \leq V_2$ ($V_2' \leq V \leq V_1'$), where $V_1 > 0$ ($V_1' < 0$). The lower boundary $V_1$ should be chosen slightly above the voltage level below which the memristor state changes are negligible, whereas $V_2$ should be set to a value not exceeding the smaller of the maximum DC voltage supplied by the circuit and the maximum safe operating voltage of the memristor (i.e., somewhat below the critical voltage at which irreversible damage to the device may occur). The negative boundary voltages $V_1'$ and $V_2'$ are selected according to the same rationale.

\noindent\rule{\linewidth}{0.4pt}
\begin{Remark}
\textcolor{black}{
In principle, for a non-volatile memristor, when  $T>T_c$, one could introduce, within the programming time, an arbitrary number of zero-input-voltage time intervals with a total duration equal to $T-T_c$, allowing a step-by-step device reconfiguration, without any %overall
difference in the state transition or in the associated energy cost, when compared to the case where the effective switching phase would happen all at once.
The idle intervals would partition the optimal voltage profile into a collection of non-consecutive traces.
It would be possible, thereby, to generate an entire family of equally optimal input voltage solutions. However, by requiring that the state–evolution function $f(x,V)$ remains different from zero with a single polarity throughout the entire $T_c$-long effective switching time interval — thus ensuring an analogue and strictly-monotonic evolution of the device state — we intentionally exclude the possibility for the optimal voltage profile to exhibit a %discontinuous,
piecewise-continuous time waveform, allowing it to exhibit one single trace throughout the state transition phase.}
%, retaining a single trace throughout the programming time.}
\end{Remark}
\noindent\rule{\linewidth}{0.4pt}

Following the optimization method, based on Pontryagin's principle, as described in~\cite{alekseev2013optimal}, the Lagrangian $L = L(x, V(x))$ for the memristor switching energy minimization problem %under focus
is defined as %\vspace{-0.3cm}

\small{
\begin{equation}
    L(x,V) = % \triangleq
    \lambda_0\frac{G_M(x)V^2}{f(x,V)}+\lambda_1\left[ \frac{1}{f(x,V)}-\frac{T}{x_f-x_i}\right],
    \label{eq:Lagrangian:1}
\end{equation}
}
\normalsize{} \newline
\noindent where $\lambda_0$ and $\lambda_1$ are Lagrange multipliers.
Considering  transitions induced by positive voltages within some allowable range $[V_1,V_2]$, if there exists a minimum for the memristor resistance switching energy $Q$ from Eq.~\eqref{eq:Q2}, then the following three conditions hold true:

\begin{enumerate}
     \item It is possible to choose two non-negative Lagrange multipliers, one of which may also be set to $0$, such that

     \begin{equation}
     \lambda_0+\lambda_1>0.
     \label{eq:cond2}\end{equation}

   % \begin{equation} \lambda_0\geq 0\;\;\textnormal{and}\;\;\lambda_1\geq 0 \label{eq:cond2}\end{equation} such that $\lambda_0+\lambda_1>0$.
    \item For each $x\in[x_i,x_f]$, it is possible to determine a particular voltage level $\hat{V}(x)\in [V_1,V_2]$, which minimizes $L$ irrespective of the constraint set by Eq.~(\ref{eq:5b}), according to
    %castable in
    the compact mathematical form

    \begin{equation} \underset{V\in[V_1,V_2]}{\textnormal{min}}\;L(x,V)=L(x,\hat{V}(x)),   \label{eq:cond1}\end{equation}

    \noindent and expressed as %, using the $\textnormal{Argmin}$ operator, %provides

    \begin{equation}
    \hat{V}(x)=\underset{V_1\leq V\leq V_2}{\textnormal{Argmin}} L(x,V). %,
    \label{eq:Argmin:1}
    \end{equation}

    %\noindent using the $\textnormal{Argmin}$ operator
    %irrespective of the constraint set by Eq.~(\ref{eq:5b}).
    \item $\lambda_1$ and $\hat{V}(x)$ satisfy the condition

   % \begin{equation} \lambda_1\cdot \beta[\hat{V}(x)]=0, \label{eq:cond3}
   % \end{equation}
    \begin{equation}
    %\lambda_1=0 \quad \text{or} \quad \beta[\hat{V}(x)]=0.
    \lambda_1 \cdot \beta[\hat{V}(x)]=0,
    \label{eq:cond3}
    \end{equation}

   \noindent which requires at least one between $\lambda_1$ and $\beta[\hat{V}(x)]$ to be identically equal to $0$.
\end{enumerate}

%We note that equation.~(\ref{eq:cond1}) can be rewritten using the $\textnormal{Argmin}$ function:
%\begin{equation}
%    \hat{V}(x)=\underset{V_1\leq V\leq V_2}{\textnormal{Argmin}} L(x,V).
%    \label{eq:Argmin:1}
%\end{equation}

In Section~\ref{sec:3}, we explore the application of the optimization method of Pontryagin's principle, expressed through the triplet of constraints (\ref{eq:cond2}), (\ref{eq:Argmin:1}), and (\ref{eq:cond3}), to the VTEAM model, which has been shown to capture rather accurately a wide class of non-volatile memristive physical realizations.

\subsection{The VTEAM model} \label{sec:2b}
\textcolor{black}{The VTEAM model~\cite{Kvatinsky2013}, developed by Kvatinsky et al. in 2015, falls in the class of first-order voltage-controlled non-volatile memristors. RESET (SET) dynamics are activated in a device if and only if a positive (negative) voltage across the devices increases above (descends below) a specific off (on) threshold voltage $v_{off}$ ($v_{on}$).
A power law governs the impact of the voltage $V$ on the switching kinetics. The VTEAM state equation is piece-wise differentiable with respect to the voltage, reading as}

\scriptsize{
\begin{numcases}{\frac{\textnormal{d} w}{\textnormal{d}t} =f(w,V)=}
      k_{off}\left(\frac{V}{v_{off}}-1\right)^{\alpha_{off}} f_{off}(w), & $0<v_{off}<V$, \label{eq:VTEAM_RESET} \\
      0, & $v_{on}\leq V\leq v_{off}$, \label{eq:VTEAM_read}\\
      k_{on}\left(\frac{V}{v_{on}}-1\right)^{\alpha_{on}} f_{on}(w), & $V<v_{on}<0$, \label{eq:VTEAM_SET}
\end{numcases}
}
\normalsize{}

\noindent where $w$ denotes the internal state variable, physically associated to a tunneling gap, while $k_{off}$, $k_{on}$, $\alpha_{off}$, $\alpha_{on}$, $v_{off}$, and $v_{on}$ are constant parameters that share a positive sign except for $k_{on}$ and $v_{on}$.
Importantly, $\alpha_{off(on)}$ is a critical parameter defining the exponent in the power law, which expresses the impact of the positive (negative) voltage $V$ on the device RESET (SET) switching kinetics.

Looking at Eqs.~(\ref{eq:VTEAM_RESET}) and (\ref{eq:VTEAM_SET}), providing the formulas for the off and on state evolution functions, respectively, $f_{off}(w)$ and $f_{on}(w)$ are window functions, that restrict the state variable $w$ to the real-valued interval $[w_{on},w_{off}]$ at all times.
In this study, we assume the mathematical form
\begin{equation}
    f_{off}(w)=\frac{w_{off}-w}{w_{off}-w_{on}}\;\;\;\text{and}\;\; f_{on}(w)=\frac{w-w_{on}}{w_{off}-w_{on}}
\end{equation}
for the window functions, ensuring $w$ to keep below its upper bound $w_{off}$ when it increases during RESET resistance switching transitions, and above its lower bound $w_{on}$ when it decreases during SET resistance switching transitions.

\textcolor{black}{We assume the memductance function $G_M(w,V)$ to be independent of the voltage $V$. The particular formula, assigned to $G_M(w)$, shows a linear relation between the device state and its conductance, reading as}
\begin{equation}
    G_M(w)=G_{max}+(G_{min}-G_{max})\frac{w-w_{on}}{w_{off}-w_{on}},\label{eq:VTEAM_memductance}
\end{equation}
where $G_{min}$ and $G_{max}$ are its minimum and maximum admissible values, respectively.

% da qui
\section{Analysis and Results} \label{sec:3}
Section \ref{sec:3a1} (\ref{sec:3a2}) derives the stimulus, which minimizes energy losses during the resistance switching transitions of a first-order memristor, the VTEAM model is fitted to, in scenarios where no (some) constraints are enforced on the allowable voltage levels, applicable across the device, and on the programming time.

\subsection{Optimal Unconstrained Solution}\label{sec:3a1}

First, we evaluate the energy costs
to be paid for turning off (on) a VTEAM memristor, while following a constant voltage-based switching control protocol, whereby
no limitation is imposed on the positive (negative) amplitude $V_0$ and duration $T_{off(on)}$ of the RESET (SET) pulse let fall across the device.

\begin{figure*}[tb]
\centering
%(a)\includegraphics[width=0.42\textwidth]{fig2a.pdf}
%(b)\includegraphics[width=0.42\textwidth]{fig2b.pdf}
(a)\includegraphics[scale=.3]{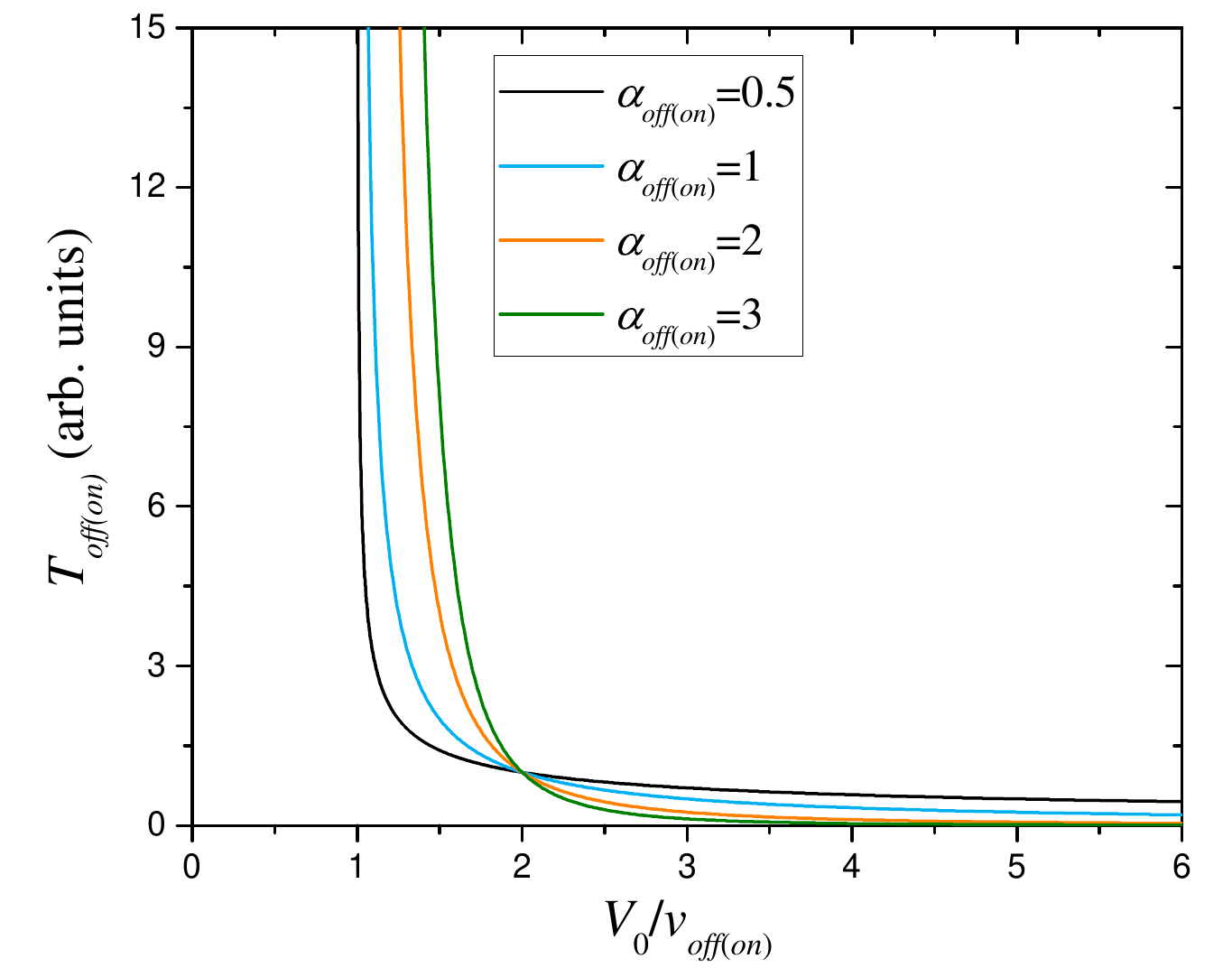}
(b)\includegraphics[scale=.3]{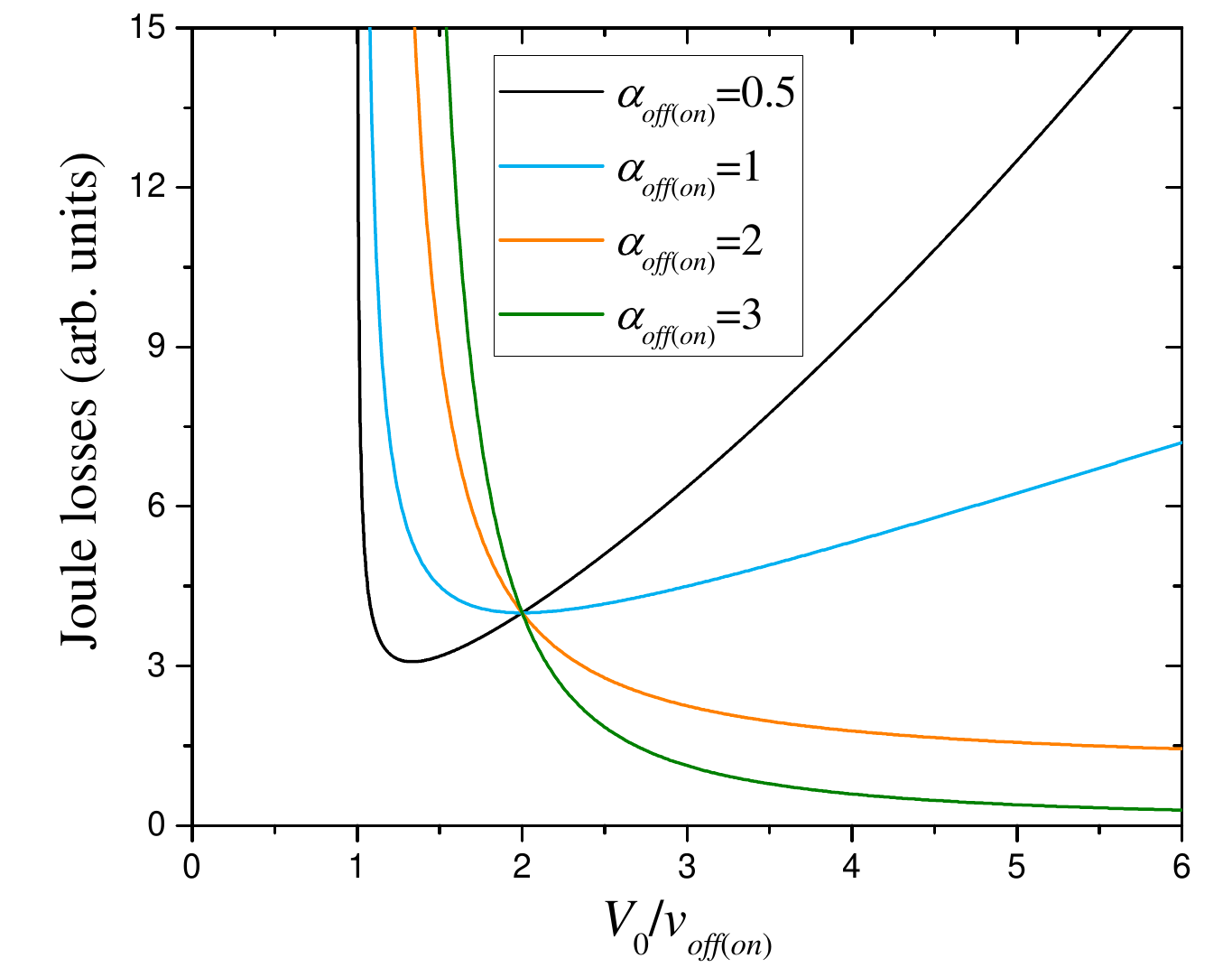}
\caption{Unconstrained control of the RESET (SET) resistance switching transition in a VTEAM memristor subjected to a square voltage pulse:
(a) programming time $T_{off(on)}$ from Eq.~(\ref{eq:Toff:VTEAM:1}), ((\ref{eq:Ton:VTEAM:1})) and (b) switching energy $Q_{off(on)}$ from Eq.~(\ref{eq:Qoff:VTEAM:1}), ((\ref{eq:Qon:VTEAM:1})) against the positive dimensionless parameter  $V_0/v_{off(on)}$
for some values, assigned to the exponent $\alpha_{off(on)}$.}
\label{fig:2}
\end{figure*}

The width $T_{off(on)}=t_f-t_i$ of the RESET (SET) voltage pulse
is found using Eq.~(\ref{eq:5a}) with $f(x,V)$ from Eq.~(\ref{eq:VTEAM_RESET}) (Eq.~(\ref{eq:VTEAM_SET})):

%can be determined by \textcolor{black}{calculating} the state integral of the reciprocal of the off (on) state evolution function $f(w,V)$, retrievable from Eq.~(\ref{eq:VTEAM_RESET}) (Eq.~(\ref{eq:VTEAM_SET}), on the left hand side of Eq.~(\ref{eq:5a}), which leads to the first (second) closed-form expression to follow:

\begin{eqnarray}
    T_{off}&=&\frac{w_{off}-w_{on}}{k_{off}\left(V_0/v_{off}-1 \right)^{\alpha_{off}}}\ln\frac{w_{off}-w_i}{w_{off}-w_f}, \label{eq:Toff:VTEAM:1} \\
    T_{on}&=&\frac{w_{off}-w_{on}}{k_{on}\left(V_0/v_{on}-1 \right)^{\alpha_{on}}}\ln\frac{w_{f}-w_{on}}{w_{i}-w_{on}},  \label{eq:Ton:VTEAM:1}
\end{eqnarray}
\noindent with $V_0>v_{off}$ in (\ref{eq:Toff:VTEAM:1})
($V_0<v_{on}$ in (\ref{eq:Ton:VTEAM:1})).
The RESET (SET) switching time $T_{off(on)}$ decreases monotonically with increases in the
strictly positive normalized pulse voltage $V_0/v_{off(on)}$, as illustrated in
Fig.~\ref{fig:2}(a). %and $T_{on}$, do not exhibit any interesting behavior.
%which is typically referred to as the voltage-time dilemma in the device physics community.
%
Furthermore, through the computation of the integral in Eq.~(\ref{eq:Q2}),
%yet without carrying out the minimization operation,
one may obtain a closed-form expression for the energy $Q_{off}$ ($Q_{on}$) necessary to increase (decrease) the internal state $w$ of the memristor from $w_i\equiv w(t_i)$ to $w_f\equiv w(t_f)$ by letting a RESET (SET) voltage pulse of positive (negative) amplitude $V_0$ fall between its terminals from $t=t_i$ to $t=t_f$, as reported in the first (second) equation to follow:  %Joule losses:

\small{
\begin{eqnarray}
    Q_{off}&=&\frac{V_0^2}{k_{off}\left(V_0/v_{off}-1 \right)^{\alpha_{off}}}\cdot
    \bigg( G_{min}(w_{off}-w_{on}) \cdot  \nonumber \\ && \ln\frac{w_{off}-w_i}{w_{off}-w_{f}}+(G_{max}-G_{min})(w_f-w_i)\bigg),\; \label{eq:Qoff:VTEAM:1} \;\;\;\;\; \\
    Q_{on}&=&\frac{V_0^2}{k_{on}\left(V_0/v_{on}-1 \right)^{\alpha_{on}}}\cdot
    \bigg( G_{max}(w_{off}-w_{on}) \cdot  \nonumber \\ &&
     \ln\frac{w_{f}-w_{on}}{w_{i}-w_{on}}+(G_{max}-G_{min})(w_i-w_f)\bigg). \label{eq:Qon:VTEAM:1} \;\;\;\;\;
\end{eqnarray}
}
\normalsize{}

%It is worth observing that
%In the unconstrained optimization,
$Q_{off(on)}$ may depend upon $V_0$ in two distinct ways depending upon the critical model parameter
$\alpha_{off(on)}$. Fig.~\ref{fig:2}(b) shows that, when $\alpha_{off(on)}$ is smaller than $2$, there exists a minimum for the locus of the RESET (SET) switching energy $Q_{off(on)}$ versus the strictly-positive scaled parameter $V_0/v_{off(on)}$.
%positive (negative) pulse amplitude $V_0$ by the positive (negative) threshold voltage  $v_{off(on)}$.
As shown in the same figure, %Fig. \ref{fig:2}(b),
if, on the other hand,
%In the second case, where
$\alpha_{off(on)}$ is equal to or larger than $2$, $Q_{off(on)}$ is a monotonically decreasing function of $V_0/v_{off(on)}$.

Now, differentiating Eq.~(\ref{eq:Qoff:VTEAM:1}) ((\ref{eq:Qon:VTEAM:1})) with respect to $V_0$ and setting the resulting expression to $0$,
the following closed-form expression is obtained for the positive (negative) RESET (SET) pulse amplitude $V^*_{off(on)}$, which minimizes the Joule losses across the memristor, %physical stack,
when
the inequality $\alpha_{off(on)}<2$ holds true:

\begin{equation}
    V^*_{off(on)} \equiv\frac{2}{2-\alpha_{off(on)}}v_{off(on)}. \label{eq:13}
\end{equation}

%The solution (\ref{eq:13}) is valid .
%circumstances,
Substituting the positive (negative) pulse height $V_{off(on)}^*$ from Eq.~(\ref{eq:13}) into Eqs.~(\ref{eq:Toff:VTEAM:1}) and (\ref{eq:Qoff:VTEAM:1}) ((\ref{eq:Ton:VTEAM:1}) and (\ref{eq:Qon:VTEAM:1})), the formulas for the resulting programming time
$T^*_{off(on)}$, coinciding with the pulse width,
and for the minimal energy $Q^*_{off(on)}$, necessary to increase (decrease) the device state from $w_i$ to $w_f$,
%necessary for the memristor to undergo a RESET (SET) transition, which results in its internal state change from $w_i$ to $w_f$,
%, in the scenario, where the pulse-driven memristor undergoes the optimal RESET (SET) transition,
%when $0<\alpha_{off(on)}<2$,
are respectively reported in the first (third) and second (fourth) expressions to follow:

\small{
\begin{eqnarray}
% \\
  T^*_{off}&=&\frac{w_{off}-w_{on}}{k_{off}}\left(\frac{2-\alpha_{off}}{\alpha_{off}}\right)^{\alpha_{off}}\ln\frac{w_{off}-w_i}{w_{off}-w_f},  \label{eq:Toff:VTEAM:2} \;\\
      Q^*_{off}&=&\frac{4v_{off}^2}{k_{off}\alpha_{off}^{\alpha_{off}}(2-\alpha_{off})^{2-\alpha_{off}}}\cdot
    \bigg( G_{min}(w_{off}-w_{on}) \cdot  \nonumber \\  && \ln\frac{w_{off}-w_i}{w_{off}-w_{f}}+(G_{max}-G_{min})(w_f-w_i)\bigg),\; \label{eq:Qoff:VTEAM:2} \\
   %  \\
%
T^*_{on}&=&\frac{w_{off}-w_{on}}{k_{on}}\left(\frac{2-\alpha_{on}}{\alpha_{on}}\right)^{\alpha_{on}}\ln\frac{w_{f}-w_{on}}{w_{i}-w_{on}}, \label{eq:Ton:VTEAM:2} \\
 Q^*_{on}&=&\frac{4v_{on}^2}{k_{off}\alpha_{on}^{\alpha_{on}}(2-\alpha_{on})^{2-\alpha_{on}}}\cdot
    \bigg( G_{max}(w_{off}-w_{on}) \cdot \nonumber \\
    &&  \ln\frac{w_{f}-w_{on}}{w_{i}-w_{on}}+(G_{max}-G_{min})(w_i-w_f)\bigg).  \label{eq:Qon:VTEAM:2}
\end{eqnarray}
}
\normalsize{}

As may be evinced from the analysis of Eq.~(\ref{eq:Qoff:VTEAM:1}) ((\ref{eq:Qon:VTEAM:1})), when $\alpha_{off(on)}\geq 2$, the Joule losses, associated to a RESET (SET) transition, are minimized by letting the pulse amplitude $V_0$
tend to positive (negative) infinity.
%(Eqs.~(\ref{eq:Qoff:VTEAM:1}) and (\ref{eq:Qon:VTEAM:1})) are  with $V_0\rightarrow \pm\infty$, respectively.
%In this limit,
If this were physically possible, the RESET (SET)
switching energy $Q_{off(on)}$ and programming time $T_{off(on)}$ would asymptotically approach zero.
%$Q^*_{off(on)}\rightarrow 0$ and $T^*_{off(on)}\rightarrow 0$.
In practice, for $\alpha_{off(on)}\geq 2$, the modulus of the positive (negative) RESET (SET) pulse amplitude $V_0$ should be set in a sub-optimal way to some level $|V_2|$, which is the smallest value between the highest voltage available in the circuit, hosting the memristor, and the maximum voltage, which can be applied across the device without jeopardizing its life expectancy.
%The latter case leads to a sub-optimal control of its OFF (ON) switching transition.
%The SET pulse amplitude should be selected in a similar way, taking into account the negative sign. %Consequently, the pulse duration would be set to the suboptimal value $T_{off(on)}(V_2)$ calculated using Equation (\ref{eq:Toff:VTEAM:1}) ((\ref{eq:Ton:VTEAM:1})) with $V_0=V_2$.

%which, in practice, yields the optimal control.

%\vspace{3mm}

%Sec.~\ref{sec:3a2}
The next section demonstrates how, when the design specifications impose limitations on the programming time $T$ and on the voltage levels applicable across the memristor, the most energetically favorable solution for the stimulus may be determined by recurring to Pontryagin's principle.
In particular, the triplet of constraints, expressed by Eqs.~(\ref{eq:cond2}), (\ref{eq:Argmin:1}), and (\ref{eq:cond3}), shall be adapted to the DAE set, composed of Eqs. (\ref{eq:VTEAM_RESET})-(\ref{eq:VTEAM_SET}) and
(\ref{eq:2}) with (\ref{eq:VTEAM_memductance}), where $w \equiv x$, and concurrently solved to derive the optimal control voltage
%concurrently.
for three different cases of the values of $\lambda_0$ and $\lambda_1$.

%the optimization problem shall be tackled by recurring to the method of Langrangian multipliers, so as to %\textcolor{red}{
%allow the  by solving the triplet of constraints, expressed by Eqs.~(\ref{eq:cond2}), (\ref{eq:Argmin:1}), and (\ref{eq:cond3}), while taking into account, concurrently, .
\subsection{Optimal Constrained Solution}\label{sec:3a2}
Without loss of generality, the analysis will focus on RESET switching transitions, assuming the minimum positive voltage $V_1$, applicable across the device, to coincide with the off %OFF
threshold %switching
voltage $v_{off}$.
Similar results may be derived \emph{mutatis mutandis} for SET switching.

{\bf Case 1: $\lambda_0=0$, $\lambda_1>0$}.
Letting $x = w$, and solving the minimization problem, expressed by the formula~(\ref{eq:Argmin:1}), where the Lagrangian $L(w,V)$ is expressed through (\ref{eq:Lagrangian:1}), in which $f(w,V)$ is replaced with its formula for $V>v_{off}$, namely (\ref{eq:VTEAM_RESET}), %-(\ref{eq:VTEAM_SET}),
and the constant term proportional to $\lambda_1$ is disregarded,
the optimal positive RESET voltage $\hat{V}(w)$, to be applied across the device for each value assumed by its internal state variable $w$ from the admissible range $[w_{i},w_{f}]$, is computed via
%by solving the optimization

\begin{equation}
    \hat{V}(w)=\underset{v_{off}\leq V\leq V_2}{\textnormal{Argmin}}  \frac{\lambda_1}{ k_{off}\left(\frac{V}{v_{off}}-1\right)^{\alpha_{off}} f_{off}(w)}.
    \label{eq:VTEAM:Vhat:1}
\end{equation}

Clearly, the largest possible value for $V$ minimizes the argument of the operator $\textnormal{Argmin}$ in Eq.~(\ref{eq:VTEAM:Vhat:1}),
irrespective of $w$. As a result,
taking also the constraint~(\ref{eq:cond3}), implying $\beta[\hat{V}(w)] = 0$, into account,
the application of our optimization procedure to case 1 simply results in the recommendation to apply the largest admissible voltage, namely
%solution of equation~(\ref{eq:VTEAM:Vhat:1}) is clearly

\begin{equation}
    \hat{V}(w)=V_2, \label{eq:VTEAM:Vhat:2}
\end{equation}

\noindent across the memristor, irrespective of its internal state $w$, over the whole programming phase temporal window $[t_i,t_f]$, whose duration $T$, computable via the state integral of the reciprocal of the off state evolution function $f(w,V)$, retrievable from Eq.~(\ref{eq:VTEAM_RESET}), in (\ref{eq:5a}),
is found to admit the expression
%depend upon initial and final internal states, referred to %as $w_i=w(t_i)$ and $w_f=w(t_f)$, respectively, according to

\begin{equation}
    %t_f
    T=\left(\frac{V_2}{v_{off}}-1\right)^{-\alpha_{off}}\int\limits_{w_i}^{w_f}\frac{\textnormal{d}w}{k_{off}f_{off}(w)}\equiv T_c,
    \label{eq:VTEAM:cond4}
\end{equation}

\noindent %\textcolor{red}{
reducing to the formula for $T_{off}(V_0)$ from Eq.~(\ref{eq:Toff:VTEAM:1}) with $V_0=V_2$.
As will be shown % discussed %shortly
in section \ref{sec:3a3}, $T_{off}(V_2)$ coincides with the shortest possible time, to be indicated as $T_{off}^{min}$,
for turning off the device, calling for the application of the highest admissible positive voltage between its terminals.
%}.
%and to coincide with the expression for $T$, once $t_i$ is chosen equal to $0$.
%
%So, we have obtained a solution in which the maximum possible voltage, $V_2$, is applied throughout the time interval $T=t_f-t_i$.
%\textcolor{red}{
Note that the optimal solution in case 1 is valid as long as the given programming time $T$
%the final time $t_f$
%of the programming phase
is precisely equal to $T_{off}(V_2)$.%, which makes this scenario  improbable in practice.

{\bf Case 2: $\lambda_0>0$, $\lambda_1=0$}.
Letting $x=w$, inserting the off state evolution function, retrievable from Eq.~(\ref{eq:VTEAM_RESET}), into the formula for the Lagrangian $L(w,V)$, obtained from (\ref{eq:Lagrangian:1}), %with $x \triangleq w$,
and assigning, without loss of generality, $\lambda_0=1$ reduces Eq.~(\ref{eq:Argmin:1}) to
%In this and the next case, we set $\lambda_0=1$.

%Using Eqs.~(\ref{eq:Argmin:1}), (\ref{eq:Lagrangian:1}), and (\ref{eq:VTEAM_RESET}-\ref{eq:VTEAM_SET}), we get
\begin{equation}
    \hat{V}(w)=\underset{v_{off}\leq V\leq V_2}{\textnormal{Argmin}}  \frac{G_M(w)V^2}{ k_{off}\left(\frac{V}{v_{off}}-1\right)^{\alpha_{off}} f_{off}(w)},
    \label{eq:VTEAM:Vhat:3}
\end{equation}

\noindent where $\hat{V}(w)$ represents the positive voltage, which, let fall across the device for each value of its state variable $w$ across the desired variation range $[w_{i},w_{f}]$, determines the most energetically favorable RESET switching transition.

%The function $\hat{V}(w)$
The argument of the Argmin operator in Eq. (\ref{eq:VTEAM:Vhat:3})
%defined by Eq.~(\ref{eq:VTEAM:Vhat:3})
depends upon $V$ %$\alpha_{off}$
in a similar manner as $Q_{off}$ in Eq.~(\ref{eq:Qoff:VTEAM:1}) is related to $V_0$ (recall also Fig.~\ref{fig:2}(b)).
Clearly, for $\alpha_{off} \geq 2$, the optimal control voltage $\hat{V}(w)$ should be set to the upper bound in its admissible range $[V_1,V_2]$, as reported in Eq.~(\ref{eq:VTEAM:Vhat:2}), irrespective of $w$.
By calculating the state integral of the reciprocal of the off state evolution function $f(w,V)$, inferable from Eq.~(\ref{eq:VTEAM_RESET}),
in (\ref{eq:5a}), the \textcolor{black}{effective} switching time $T_c$, here smaller than or equal to the programming time $T$, when $\beta[\hat{V}(w)]$ is respectively different from or equal to $0$, in view of %the constraint
(\ref{eq:cond3}),
%or $\beta[\hat{V}(w)]= 0$, respectively,
is found to admit
%\textcolor{red}{
the same closed-form expression as $T_{off}(V_0)$ in Eq.~(\ref{eq:Toff:VTEAM:1}) for $V_0=V_2$. %}. %, upon setting $t_i$ to $0$}.
%that should be met.

If, on the other hand, %consider
$\alpha_{off}<2$, %$0<\alpha_{off}<2$.
the solution to the minimization problem, defined via Eq.~(\ref{eq:VTEAM:Vhat:3}), is % it follows that
\begin{equation}
\hat{V}(w) = \textnormal{min}\left\{ V_{off}^*,V_2\right\}.
  %  \hat{V}(w) = \textnormal{min}\left\{ \frac{2}{2-\alpha_{off}}v_{off},V_2\right\}.
     \label{eq:VTEAM:Vhat:4}
\end{equation}
Once again, calculating the state integral of the reciprocal of the off state evolution function in %condition
(\ref{eq:5a}), allows to compute the \textcolor{black}{effective} switching time $T_c$, which, as explained above, is smaller than or equal to the programming time $T$.
%precedes the programming phase final time $t_f$.   %should be met.
%\textcolor{red}{ %For $t_i=0$,
If %$\frac{2}{2-\alpha_{off}}v_{off}<V_2$,
%$2/(2-\alpha_{off})\cdot v_{off}<V_2$,
$V_{off}^*<V_2$,
the expression for $T_c$ reduces to the formula for $T_{off}^*$ in Eq.~\eqref{eq:Toff:VTEAM:2}, else
to the formula for $T_{off}(V_0)$ in Eq.~(\ref{eq:Toff:VTEAM:1}) with $V_0=V_2$.
%(as will be discussed in Sec.~\ref{sec:3a3}, $T_{off}(V_2)$ coincides with the shortest time, indicated as $T_{off}^{min}$, necessary for the device to reset when the maximum possible voltage is applied between its terminals).
%}

All in all, in case 2, the control voltage $\hat{V}(w)$, inducing the most energetically-favorable RESET switching transition across the device, is given by Eq.~(\ref{eq:VTEAM:Vhat:2}) ((\ref{eq:VTEAM:Vhat:4})) for values of the parameter $\alpha_{off}$ larger than or equal to $2$ (smaller than $2$), and should be applied across the device over a $T_c$-long time interval $[t_i,t_s]$,
with $t_s\leq t_f$. In particular, when $t_s<t_f$, no voltage should further stimulate the device in the remainder of the programming phase, i.e. for $t\in\{ t_s,t_f]$. \\
%in the remainder of the programming phase.  \\
%, where, setting $t_i$ to $0$, $t_c$ is smaller than $T$. \\
%
%depending on the value of $\alpha_{off}$.
%Moreover, the switching time should satisfy $t_c\leq T$.

{\bf Case 3: $\lambda_0>0$, $\lambda_1>0$}.
In this case, letting $x=w$, and replacing $f(w,V)$ with the formula it admits for $V>v_{off}$, specifically (\ref{eq:VTEAM_RESET}), inside the expression for the Lagrangian $L(w,V)$, expressed by Eq.~(\ref{eq:Lagrangian:1}), % for $x=w$,
the control voltage $\hat{V}(w)$, to be applied across the device for reducing as much as possible the Joule losses, associated with an increase of its internal state from $w_i$ to $w_f$, may be computed via the minimization problem
%is given by

\begin{equation}
    \hat{V}(w)=\underset{v_{off}\leq V\leq V_2}{\textnormal{Argmin}}  \frac{G_M(w)V^2+\lambda_1}{ k_{off}\left(\frac{V}{v_{off}}-1\right)^{\alpha_{off}} f_{off}(w)},
    \label{eq:VTEAM:Vhat:5}
\end{equation}

\noindent where $\lambda_0$ is arbitrarily chosen equal to $1$.
%, while,
%
The formula~(\ref{eq:5a}) for the effective switching time
now reduces to %boiling down to

\begin{equation}
    T_c=\int\limits_{w_i}^{w_f}\frac{\textnormal{d}w}{k_{off}\left(\frac{\hat{V}(w)}{v_{off}}-1\right)^{\alpha_{off}}f_{off}(w)},
    \label{eq:VTEAM:cond5}
\end{equation}
\noindent
where $\hat{V}(w)$ is the optimal control voltage expressed via \eqref{eq:VTEAM:Vhat:5}.
In view of %the constraint
Eq.~(\ref{eq:cond3}), implying $\beta[\hat{V}(w)]=0$, the \textcolor{black}{effective} switching time $T_c$ coincides here with the pre-defined programming time $T$. Its knowledge permits then to use \eqref{eq:VTEAM:cond5} to choose a suitable value for $\lambda_1$, in case the formula for the optimal control voltage $\hat{V}(w)$ finally includes it.
\textcolor{black}{In these situations, increasing progressively $\lambda_1$ from $0$, and computing the state integral on the right-hand side of Eq. (\ref{eq:VTEAM:cond5}) for each value this parameter assumes during its forward sweep, allows to determine a function providing the dependence of the effective switching time $T_c$ upon this Lagrange multiplier. The value, chosen for $\lambda_1$ falls then for the solution to the equality $T_c(\lambda)=T$.}

% , if necessary
%if necessary, $\lambda_1$ should be selected, on the basis of the optimal solution $\hat{V}(w)$ itself, through the use of

%This knowledge allows to use
%which is adapted from ,
%expressing the \textcolor{black}{effective} switching time $T_c$, known and equal to the pre-defined programming
%phase final
% time $T$,

%and further coincides with $T$ for $t_i=0$.
%in terms of the state integral of the reciprocal of the state evolution function
%\noindent , and
Once again two scenarios are possible.
If $\alpha_{off}\geq 2$, the control voltage $\hat{V}(w)$, solving the minimization problem (\ref{eq:VTEAM:Vhat:5}), is the upper bound in its admissible range, as reported in Eq.~(\ref{eq:VTEAM:Vhat:2}).
In these circumstances, the formula ~(\ref{eq:VTEAM:cond5}) for the \textcolor{black}{effective} switching time $T_c$ boils down to the expression reported in Eq.~(\ref{eq:VTEAM:cond4}), %\textcolor{red}{
reducing to the formula (\ref{eq:Toff:VTEAM:1}) for $T_{off}(V_0)$ with $V_0=V_2$, %}, %, upon setting $t_i=0$},
i.e. to $T_{off}^{min}$, %(see section \ref{sec:3a3}),
which makes a selection for $\lambda_1$ unnecessary.
%Moreover, the condition~(\ref{eq:VTEAM:cond5}), which takes the form of Eq.~(\ref{eq:VTEAM:cond4}), must be satisfied exactly.

When on the other hand $\alpha_{off}<2$, the minimization problem (\ref{eq:VTEAM:Vhat:5}) reduces to %we arrive to the equation

\begin{equation}
    (2-\alpha_{off})V^2-2v_{off}V-\frac{\alpha_{off}\lambda_1}{G_M(w)}=0.
    \label{2nd_order_poly}
\end{equation}

\noindent Disregarding the negative root,
%of the second-order polynomial (\eqref{2nd_order_poly})
%associated to the choice of the negative sign in the formula for its general %admissible
%solution, %to this second-order polynomial,
which does not satisfy the inequality $\hat{V}>v_{off}$,
the only admissible zero for the second-order polynomial (\ref{2nd_order_poly}),
%the only possible optimal voltage,
indicated as $\tilde{V}(w)$,
%and associated to the choice of the positive sign in the general formula for the roots of the second-order polynomial,
is found to depend upon the device internal state $w$ via
%whose solution (the minus sign solution is disregarded, as it does not satisfy the condition $V>v_{off}$) is

\begin{equation}
    \tilde{V}(w)=\frac{v_{off}+\sqrt{v_{off}^2+\lambda_1\alpha_{off}(2-\alpha_{off})/G_M(w)}}{2-\alpha_{off}}.
     \label{eq:VTEAM:Vhat:6}
\end{equation}

%\noindent which satisfies \textcolor{red}{the condition %inequality
%
%\begin{equation}
%\tilde{V}(w)>2/(2-\alpha_{off})v_{off},
%\label{ineq}
%\end{equation}
%
%\noindent irrespective of $w$, a constraint which will be used later on in section \ref{sec:4}.}
%All in all,
\noindent Thus, for $\alpha_{off}< 2$, the optimal control voltage $\hat{V}(w)$ may be computed via

 \begin{equation}
    \hat{V}(w)=\textnormal{min}\left\{\tilde{V}(w),V_2\right\}.
     \label{eq:VTEAM:Vhat:7}
\end{equation}

%\noindent while $\lambda_1$ is defined by Eq.~(\ref{eq:VTEAM:cond5}).
%\textcolor{red}{
The formula for the \textcolor{black}{effective switching} time %interval
$T_c$, coinciding with the programming time $T$,
%here equal to the %pre-defined
%programming time interval $T$, %}
%final time $t_f$,
is %computed via
given by Eq.~(\ref{eq:VTEAM:cond5}), where $\hat{V}(w)$ is replaced by the expression in (\ref{eq:VTEAM:Vhat:7}).
%With $t_i=0$, the final formula for $t_c$ also coincides with $T$.
%Since $T$ is %typically
%part of the design specifications, the formula for $T_c$ may be used to determine a suitable value for $\lambda_1$.
This allows to determine a suitable value for $\lambda_1$. This may always be achieved via numerical methods, while an analytical treatment may pose challenges in some scenarios, especially when $\tilde{V}(w)$ from Eq.~ (\ref{eq:VTEAM:Vhat:6}) defines the optimal control voltage $\hat{V}(w)$ from Eq.~(\ref{eq:VTEAM:Vhat:7}) for some values of $w$, only. \textcolor{black}{In any case, once the optimal voltage stimulus $\hat{V}(w)$ is identified, equation (\ref{eq:5}), with $x=w$ and $V=\hat{V}(w)$, is used to derive a formula for the time $t$ as a function of the state $w$. This allows to find the temporal course of the optimal control voltage $\hat{V}(t)$ parametrically from the knowledge of $\hat{V}(w)$ and of $t(w)$ so as to induce the desired state transition from $w_i$ to $w_f$.}         \\
All in all, in case 3, the optimal control voltage $\hat{V}(w)$ should be supplied across the entire programming time according to Eq.~(\ref{eq:VTEAM:Vhat:2}) ((\ref{eq:VTEAM:Vhat:7})) for
$\alpha_{off}\geq2$ ($\alpha_{off}<2$).
As a final remark, the first of these two scenarios may only occur
when $T=T_{off}^{min}$. %, which makes its appearance improbable.
%As a final remark, while in the first case there exists one and only one possible value for $T_c$, namely $T_{off}^{min}$,
%(see section \ref{sec:3a3}),
%in the second case $T_c$ assumes the same value specified for $T$. \\
%the same values as $T$ does. \\

%In the last case of parameters $\lambda_0$ and $\lambda_1$, the control voltage is supplied throughout the time interval $[0,T]$ and is either given by Eq.~(\ref{eq:VTEAM:Vhat:7}) or Eq.~(\ref{eq:VTEAM:Vhat:2}) depending on $\alpha_{off}$. We note that
%the value of $\lambda_1$ is derived from Eq.~(\ref{eq:VTEAM:cond5}) with $\hat{V}(w)$ given by (\ref{eq:VTEAM:Vhat:7}), but analytically solving this equation is not always simple.
%In particular, the solution to Eq.~(\ref{eq:VTEAM:cond5}) poses challenges when Eq.~(\ref{eq:VTEAM:Vhat:6}) is the optimal voltage only in a limited interval of $w$. However, $\lambda_1$ can still be computed using numerical methods in such scenarios.

\noindent\rule{\linewidth}{0.4pt}
\begin{Remark}
\textcolor{black}{
Analysing the solution to the optimization problem in case 3, for $\alpha_{off}<2$, i.e. Eq.~(\ref{2nd_order_poly}),
the optimal RESET switching voltage $\hat{V}(w)$ and the conductance  $G_M(w)$ are found to satisfy the following \emph{invariant of motion}, provided $V_2> \tilde{V}(w)$: %It is interesting to note that, for the VTEAM model with $\alpha_{off}<2$,
%, provided
%that
%the highest possible voltage level is so large to satisfy the inequality
%$V_2\geq \tilde{V}(w)$, implying the highest possible voltage $V_2$, applicable across the device, is sufficiently large:}
 \begin{equation}
 G(w)\hat{V}(w)\left(\hat{V}(w)-V_{off}^*\right)=\text{const},
%G(w)\hat{V}(w)\bigg(\hat{V}(w)-\frac{2v_{off}}{2-\alpha_{off}}\bigg)=\text{const},
   \label{eq:VTEAM:Invariant}
\end{equation}
}
\textcolor{black}{\noindent
where $\text{const}$ denotes a constant proportional to the Lagrange multiplier $\lambda_1$.}

\textcolor{black}{
Furthermore, when this constant is zero, the existence of the invariant of motion, defined through Eq. (\ref{eq:VTEAM:Invariant}), applies to the case, where the memristor would be excited by an optimal sequence of two rectangular pulses, the first (second) of which holding a height equal to the discrete level
%$\hat{V}=V_{off}^{*}=2/(2-\alpha_{off})\cdot v_{off}$
$\hat{V}=V_{off}^{*}$
($\hat{V}=0$), and stretching out across the $T_c$-long ($(T-T_c)$-long) time interval $[t_i,t_s]$ ($[t_s,t_f]$), %. , with $T_c=t_s-t_i \leq T=t_f-t_i$,
as specified in case 2. }

\textcolor{black}{
All in all, allowing non-negative constants, the invariant of motion
~(\ref{eq:VTEAM:Invariant}) encompasses all the most energetically-favorable solutions, which our optimization strategy identified for the RESET switching voltage of the memristor, the VTEAM model is fitted to, for $\alpha_{off}<2$, under the situation the maximum admissible voltage level $V_2$ does not limit the forward excursion of the optimal programming voltage.}

\textcolor{black}{
Finally, in the special case, where $\alpha_{off}=1$, in the limit for $v_{off}\to 0$, $k_{off}\to 0$, and $k_{off}/v_{off}\to \text{const}'$, with $\text{const}'$ expressing an arbitrary non-zero constant, the invariant of motion \eqref{eq:VTEAM:Invariant} reduces exactly to the constant-power law, first identified in \cite{slipko2024reduction}, for the ideal memristor:}
\begin{equation}
G_M(w)\hat{V}^2(w)=\text{const},
   \label{eq:VTEAM:PowerLaw}
\end{equation}\textcolor{black}{
\noindent which allows to identify the positive voltage stimulus, which allows to reset in the least power-hungry form any first-order memristive system, whose off switching rate is proportional to the applied %positive
voltage, i.e. where the off state evolution function is of the form $f(x,V)\sim V$ -- see Eq.~(\ref{eq:VTEAM_RESET}) for the VTEAM model -- in the unconstrained case, i.e. when no limitation restricts the  dynamic range of the %positive
programming input. }

\textcolor{black}{
 An invariant of motion could be similarly derived \emph{mutatis mutandis} for the optimal off-to-on resistance switching transition as well.}
\end{Remark}
\noindent\rule{\linewidth}{0.4pt}

\begin{figure*}[tb]
\centering
%(a)\includegraphics[width=0.42\textwidth]{fig3a.pdf}
%(b)\includegraphics[width=0.42\textwidth]{fig3b.pdf}
(a)\includegraphics[scale=.3]{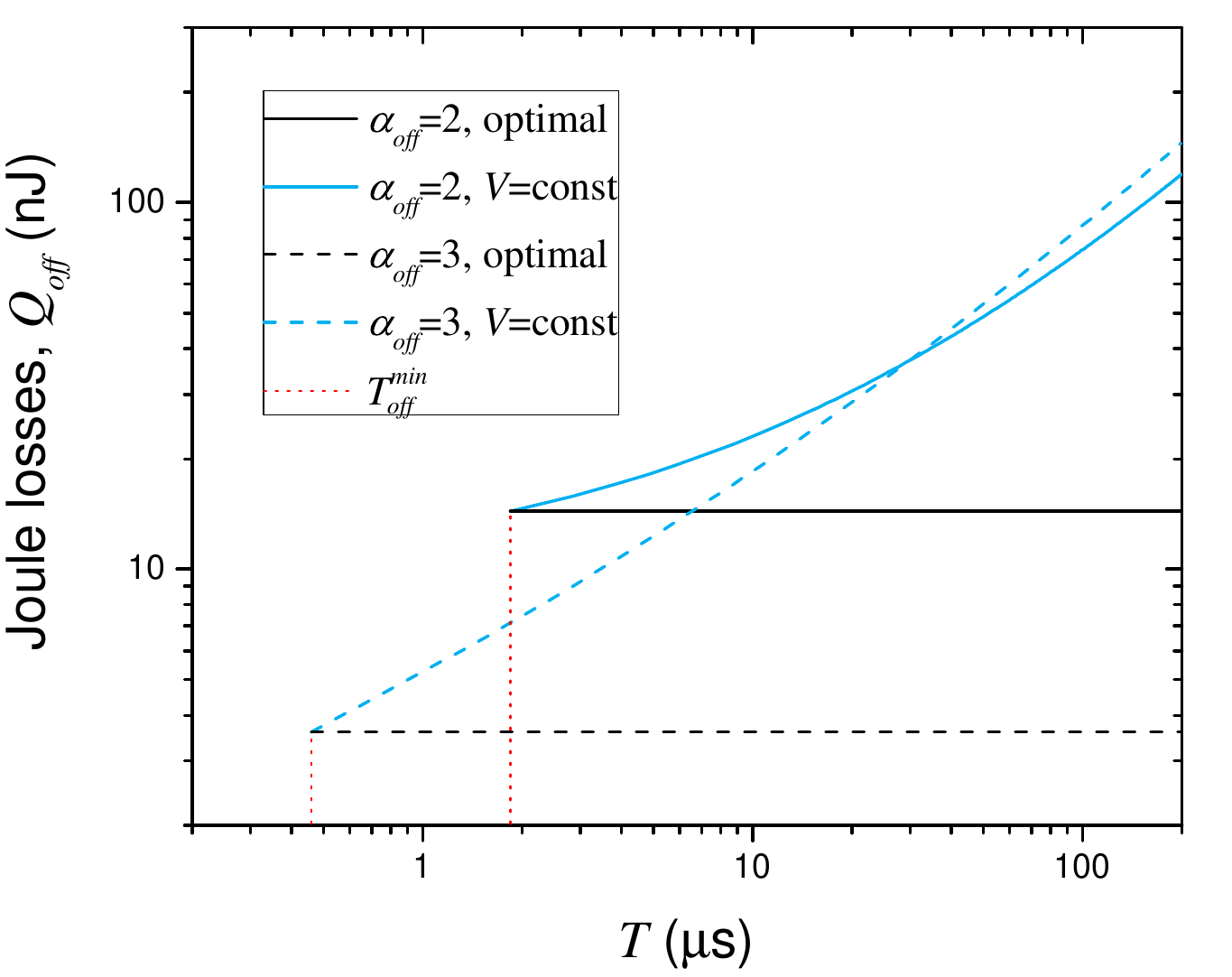}
(b)\includegraphics[scale=.3]{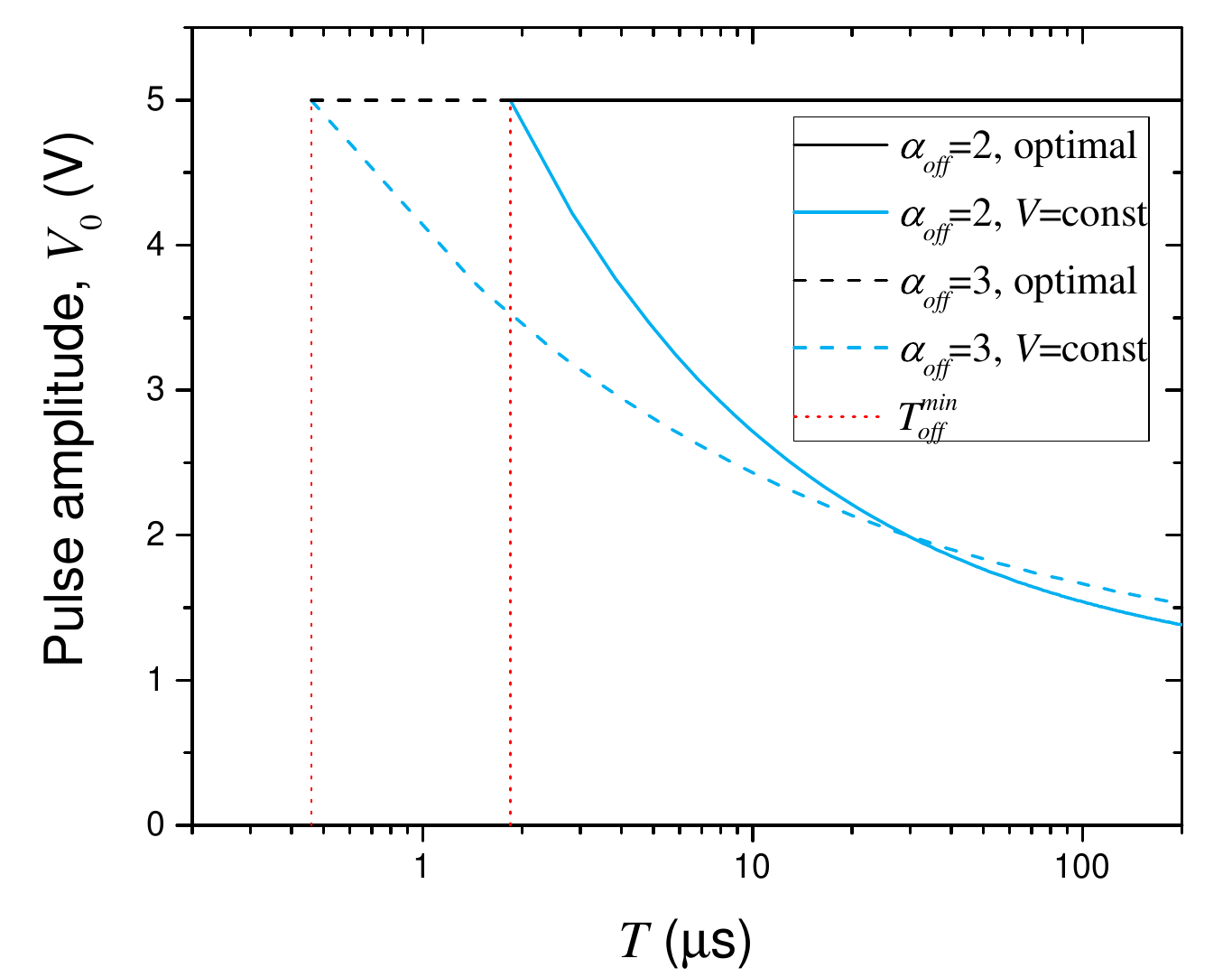}
\caption{
Comparison between the RESET energy costs incurred during the application of the constant voltage-based and optimal voltage-based switching strategies to a  VTEAM memristor, for a couple of values, specifically $2$ and $3$, assigned to the parameter $\alpha_{off}$. %\in\{2,3\} \geq 2$.
(a) RESET switching energy $Q_{off}$ and (b) pulse amplitude $V_0$ against programming time $T$. % programming phase temporal interval $T$.
The VTEAM model parameter setting, employed in these investigations, reads as follows: $w_i=0.1$, $w_f=0.9$, $w_{on}=0$, $w_{off}=1$, $v_{off}=1$~V, $k_{off}=10^5$~s$^{-1}$, $G_{min}=10^{-5}$~S, $G_{max}=10^{-3}$~S, $V_1=1$~V, $V_2=5$~V. $T^{min}_{off}$ is found to be equal to $1.84$~{\textmu}s for $\alpha_{off}=2$ and to $0.460$~{\textmu}s for $\alpha_{off}=3$.}
%As a result, $T^{min}_{off}$ is found to be equal to $1.84$~{\textmu}s for $\alpha_{off}=2$ and to $0.460$~{\textmu}s for $\alpha_{off}=3$.}
\label{fig:8}
\end{figure*}

\begin{figure*}[tb]
\centering
%(a)\includegraphics[width=0.42\textwidth]{fig4a.pdf}
%(b)\includegraphics[width=0.42\textwidth]{fig4b.pdf}\\
%(c)\includegraphics[width=0.42\textwidth]{fig4c.pdf}
%(d)\includegraphics[width=0.42\textwidth]{fig4d.pdf}
(a)\includegraphics[scale=0.31]{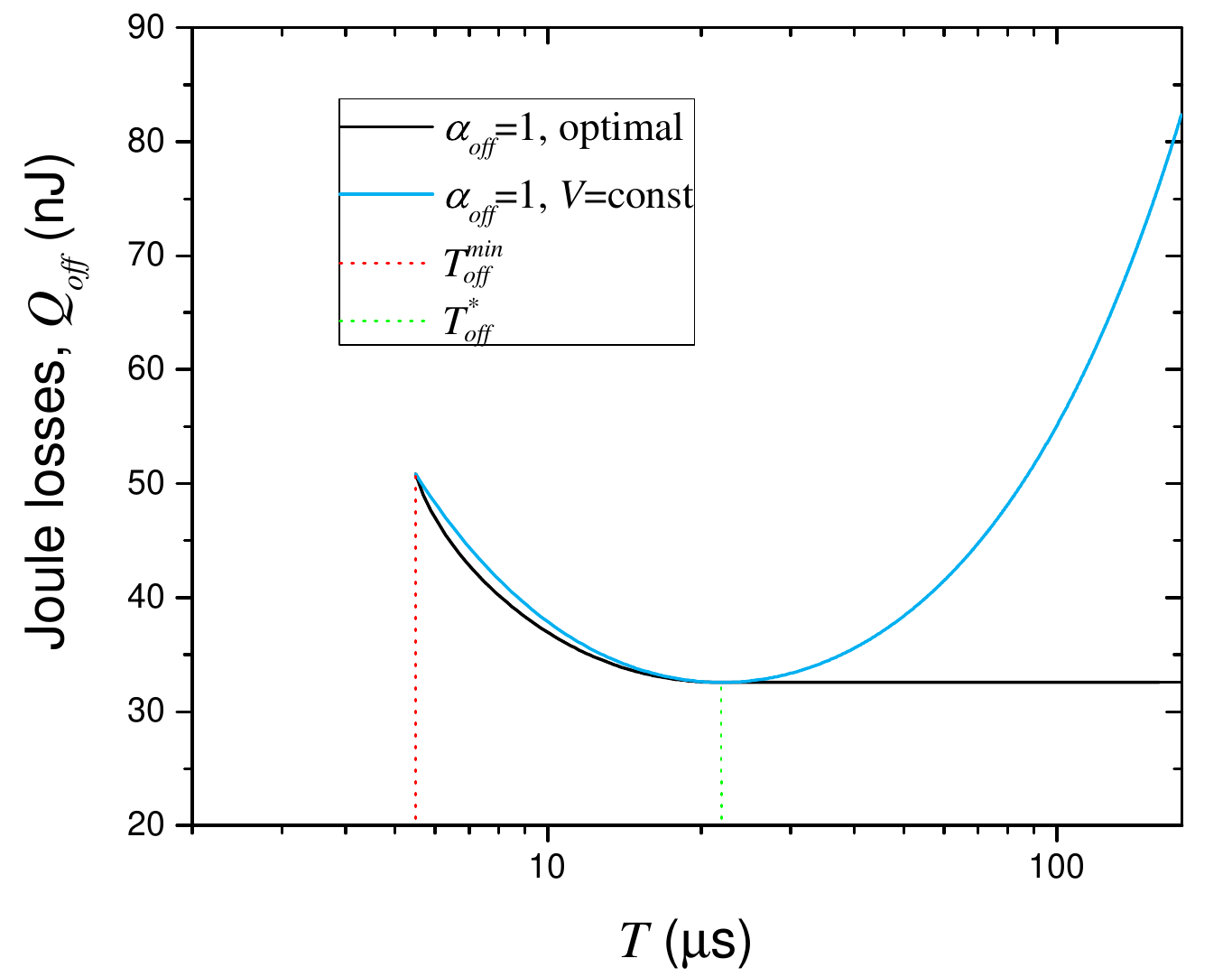}
(b)\includegraphics[scale=0.31]{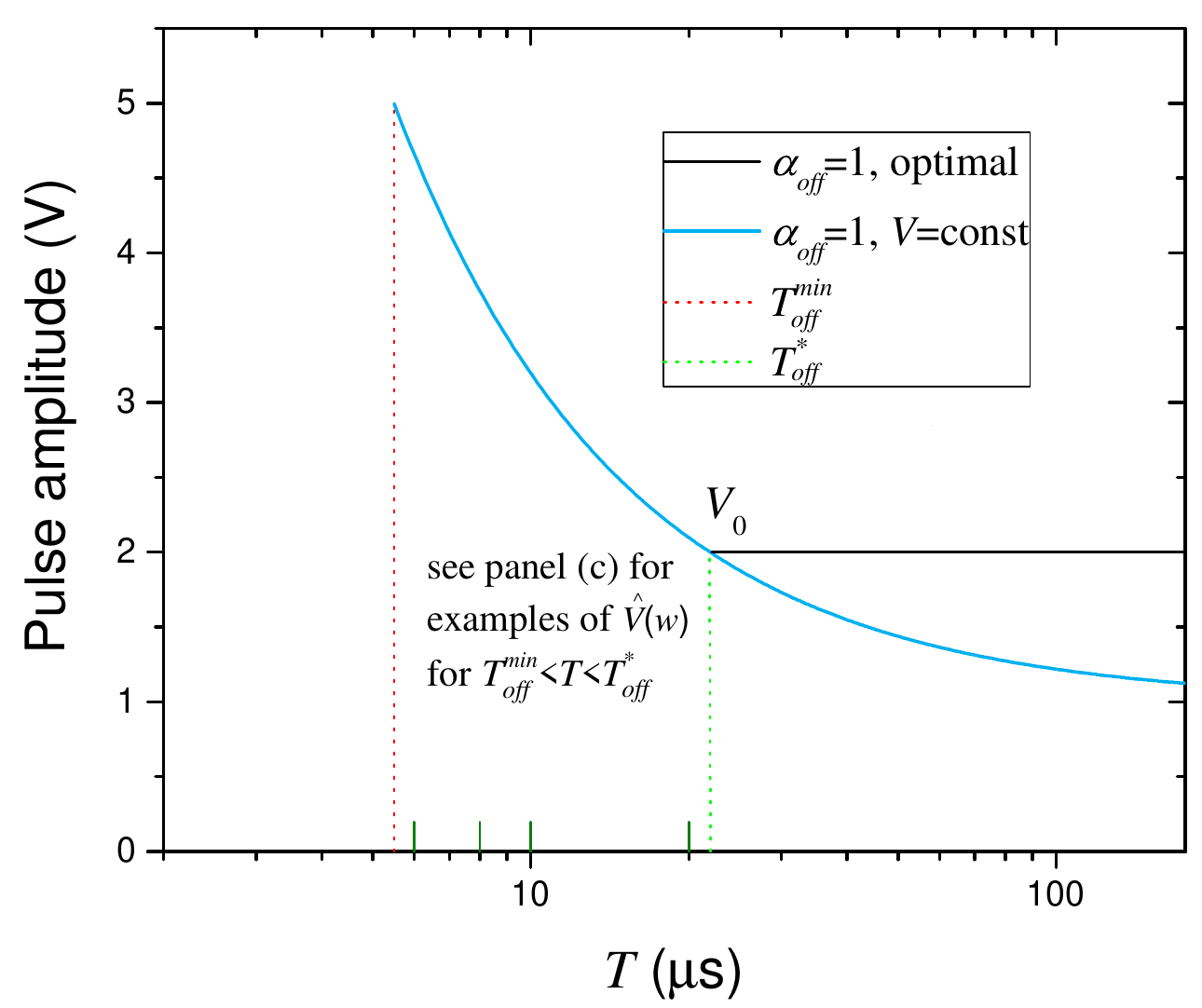}\\
(c)\includegraphics[scale=0.31]{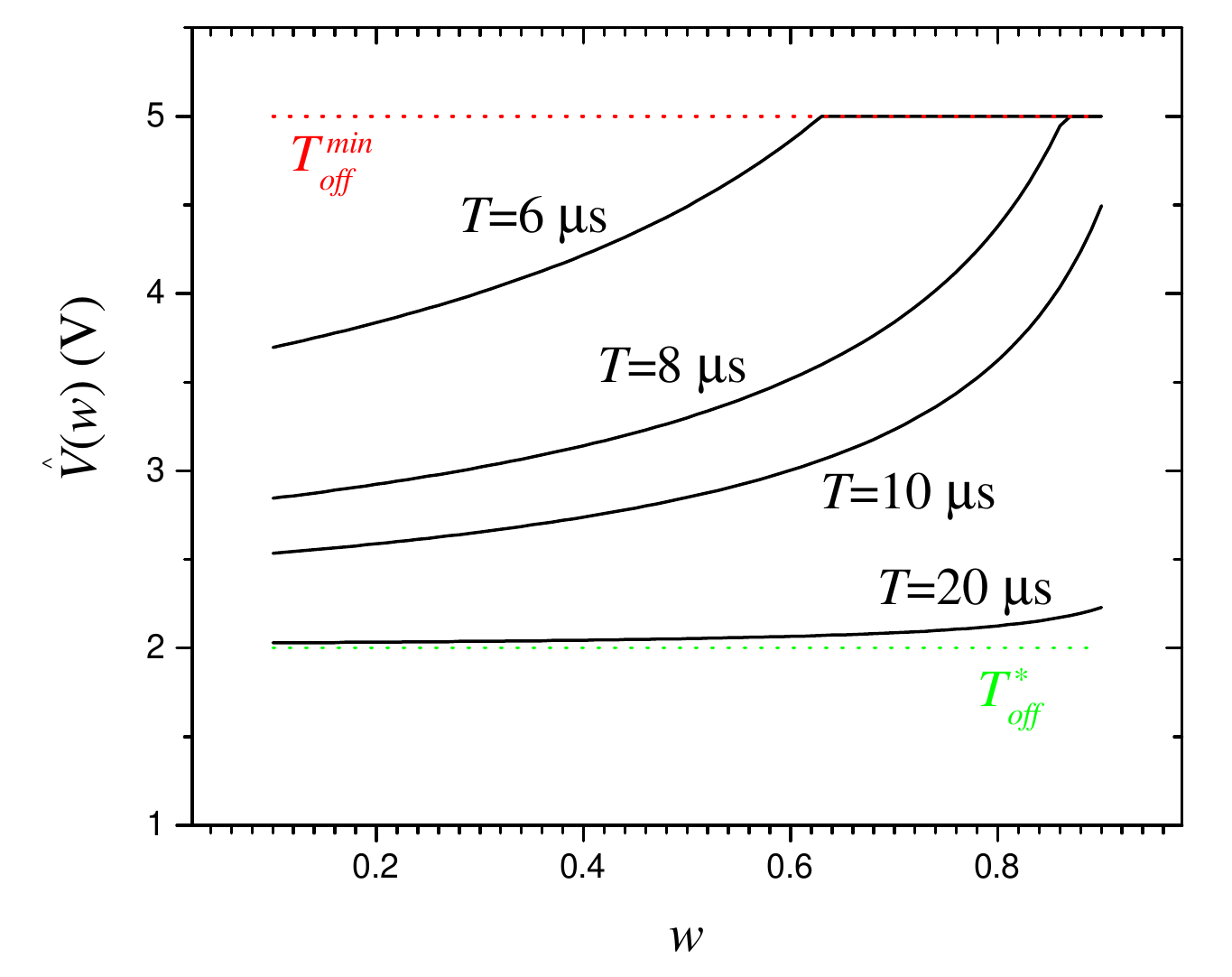}
(d)\includegraphics[scale=0.31]{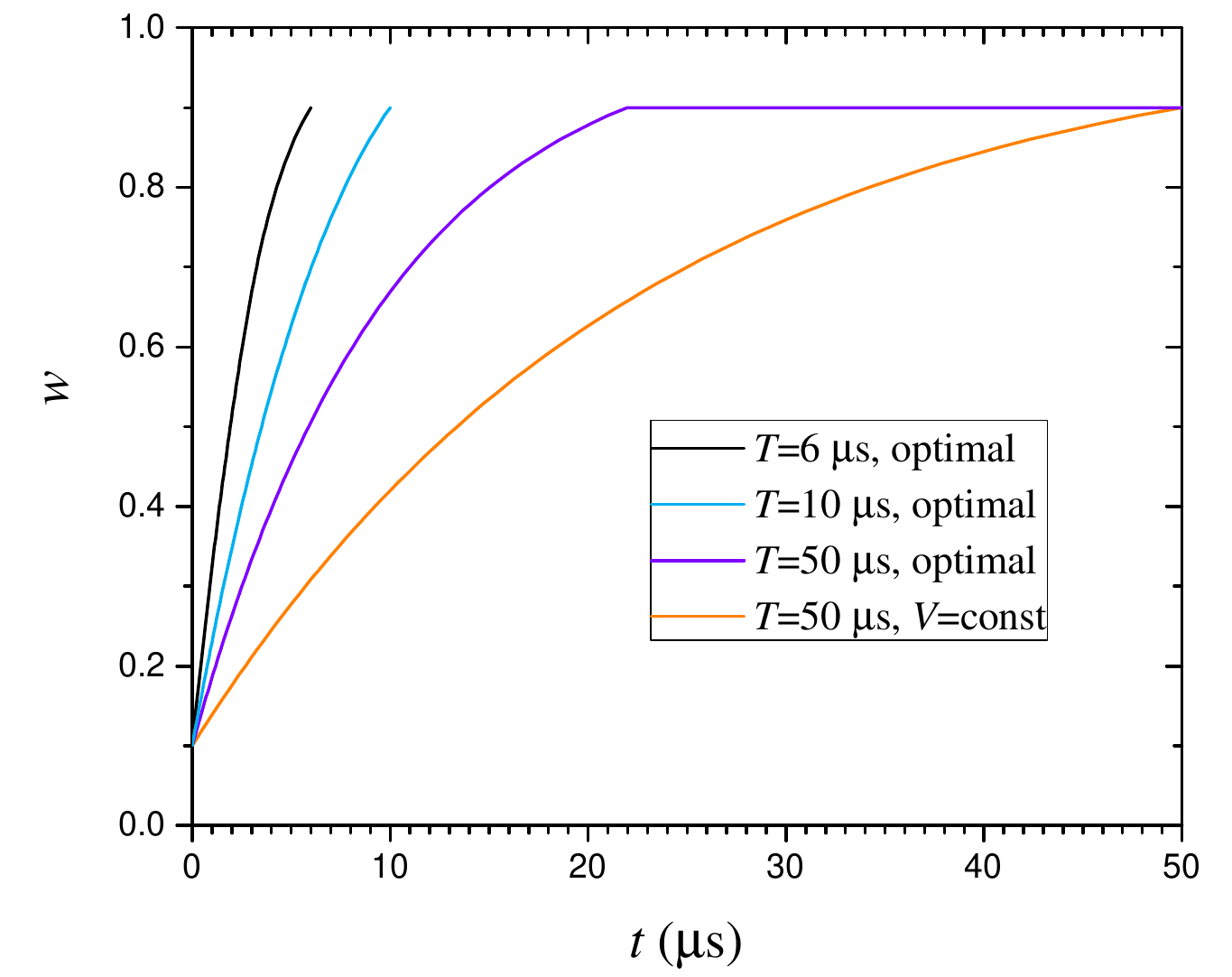}
\caption{\textcolor{black}{Comparison between the RESET energy costs due to the application of the constant voltage-based and optimal voltage-based control paradigms to a VTEAM memristor for $\alpha_{off}=1$.}
(a) RESET switching energy $Q_{off}$ and (b) pulse amplitude $V_0$ against programming time $T$. %programming phase temporal interval $T$.
%da qui
\textcolor{black}{For the optimal switching control protocol, only the case $T>T_{off}^*$ is visualized in plot (b). To avoid clutter, in fact, the case $T_{off}^{min}<T<T_{off}^*$ is accounted for in plot (c), showing the dependence of the optimal control voltage $\hat{V}(w)$ from Eq.~(\ref{eq:VTEAM:Vhat:7}) upon the device state $w$ for a number of values assigned to the programming time.
The values for $T$, labeling the loci in (c),
are indicated in (b) as short vertical lines crossing the horizontal axis.
(d) Time course of the memristor %internal
state $w$ upon assigning a couple of values (one value) from the interval $T_{off}^{min}<T<T_{off}^{*}$ ($T>T_{off}^{*}$), specifically $6$~{\textmu}s and $8$~{\textmu}s ($50$~{\textmu}s), to the programming time %phase temporal width
$T$, according to the optimal voltage-based switching
control protocol. The time evolution of $w$ under the constant voltage-based switching control protocol is also shown for one case, i.e. when $T=50$~{\textmu}s. Except for $\alpha_{off}$, set here to $1$, the very same parameter setting as reported in the caption of Fig. \ref{fig:8} was assumed for the numerical analysis this plot illustrates. Notably, here $V_{off}^{*}=2$~V, $V_2=5$~V, $T_{off}^{min}=5.493$~{\textmu}s,
and $T_{off}^{*}=21.972$~{\textmu}s.
} }
\label{fig:9}
\end{figure*}

Fig.~\ref{fig:8} (\ref{fig:9}) compares the Joule losses across the device, while undergoing RESET switching transitions under constant voltage-based and optimal voltage-based control, respectively,
for scenarios where the parameter $\alpha_{off}$ is larger than or equal to $2$ (smaller than $2$).
%, respectively.
In the constant voltage-based switching control protocol, a square voltage pulse of appropriate positive height $V_0$ is applied during the entire %allowable
programming phase temporal window $[t_i,t_f]$.
%Setting $t_i$ to $0$, which makes $T\equiv t_f$,
The expression for $V_0$ as a function of the programming time %phase temporal interval
$T=t_f-t_i$,
%equivalent to its upper bound $t_f$ upon setting $t_i$ to $0$,
is retrievable from Eq.~(\ref{eq:Toff:VTEAM:1}) with $T_{off}=T$, reducing to

\begin{equation}
    V_0=v_{off}\left(1+ \left( \frac{(w_{off}-w_{on})\ln\frac{w_i-w_{off}}{w_f-w_{off}}}{k_{off}T} \right)^\frac{1}{\alpha_{off}}\right).
    \label{constant_control_voltage}
\end{equation}

Substituting (\ref{constant_control_voltage}) into the formula for $Q_{off}$ from Eq.~(\ref{eq:Qoff:VTEAM:1}) allows to estimate the Joule losses in the device during the application of the constant voltage-based paradigm for inducing RESET transitions across its physical medium.
Inspecting both Figs.~\ref{fig:8} and ~\ref{fig:9}, obtained upon the very same %specific
model parameter setting, as reported in the caption of the first one,  %illustrations,
except for the choice of the critical coefficient $\alpha_{off}$,
it becomes apparent that, irrespective
%irrespective of the decision to
%whether the
%irrespective of the form of
%select the voltage stimulus %is selected
%on the basis of a constant or optimal
of the control paradigm, no solution exists if the value specified for the programming time %phase temporal interval
$T$ is smaller than the shortest possible RESET programming
%off switching
time, indicated as $T_{off}^{min}$, and defined in the sub-section to follow.

\subsection{Shortest RESET (SET) programming time}\label{sec:3a3}
%It is important to understand that
There are situations, in which limitations in the operating principles of a voltage-controlled memristor, %device, %corresponding to certain parameter range restrictions in the corresponding model,
captured through some predictive DAE set,
%used to model its nonlinear dynamics,
prevent a certain voltage stimulus, applied between its terminals, from switching it successfully from one resistance state to another one.
%switching transition across the corresponding
%completely block the intended switching.
For example, with reference to the VTEAM model, if the minimum duration $T_{off(on)}$ of a programming voltage pulse, which is capable to induce a successful increase (decrease) in the internal state $w$ from $w_i$ to $w_f$, when, endowed with the most positive (most negative) allowable height $V_2$, is let fall across the device,
%as a positive (negative) voltage %\textcolor{red}{
%$+(-)V_2$ %} of the highest admissible modulus is let fall across the device,
%$V_2$ of the largest admissible modulus %with the largest admissible modulus
%is let fall across the memristive physical medium, what we refer to as \emph{shortest off (on) programming time}, and indicate as $T_{off(on)}^{min}$,
is larger than the programming time
%phase time interval
$T$, assigned beforehand as a design specification, the off (on) switching transition will not produce the intended result.
Either a longer pulse or a higher pulse would be necessary to achieve the desired purpose.
This is the reason why Figs.~\ref{fig:8} and ~\ref{fig:9} reported no solution to the Joule loss minimization problem for off switching under the hypothesis $T < T_{off}^{min}$.
 % only.

$T_{off(on)}^{min}$ is referred to as the shortest RESET (SET) programming time.
%the memristive device cannot be switched to the final state $w_f$, as a longer pulse or a higher pulse amplitude would be necessary.
% da qui
A closed-form expression for the RESET (SET) programming time
%off (on) switching time
$T_{off(on)}$ as a function of the positive (negative) pulse amplitude $V_0$ for off (on) switching
%RESET (SET) transition
%scenarios
was reported in Eq.~(\ref{eq:Toff:VTEAM:1}) ((\ref{eq:Ton:VTEAM:1})) within section \ref{sec:3a2}.
 %respectively).
%In the case of the VTEAM model (and, in principle, majority of other models),
%Let us define the shortest switching time as the minimum time necessary to induce the motion of the internal state $w$ from $w_i$ to $w_f$ as the maximum admissible voltage, here $V_2$, is applied across the device.
%The shortest RESET (SET) %switching programming time depends critically upon the parameter $\alpha_{off(on)}$.
Focusing on the off switching case, using $V_0=V_2$,
%device on-to-off transitions, %in particular, we define
%transition,
the shortest programming time
%switching time
$T^{min}_{off}$ is defined as

\begin{equation}
    T^{min}_{off} = \frac{w_{off}-w_{on}}{k_{off}\left(V_2/v_{off}-1 \right)^{\alpha_{off}}}\ln\frac{w_{off}-w_i}{w_{off}-w_f}. \label{eq:Tminoff}
\end{equation}

%as reported in the captions of
%in the scenarios illustrated in
%several critical times related to the examples presented in
%Figs.~\ref{fig:8} and ~\ref{fig:9}, %are listed here:
 %$T^{min}_{off}=0.460$~{\textmu}s for $\alpha_{off}=3$, $T^{min}_{off}=1.84$~{\textmu}s ($\alpha_{off}=2$), $T^{min}_{off}=5.49$~{\textmu}s
 %and $T^*_{off}=22.0$~{\textmu}s
 %for $\alpha_{off}=1$.
 %These times are shown by
 %where their location along the horizontal axis is
 %indicated through red dashed vertical lines. %in Figs.~\ref{fig:8} and ~\ref{fig:9}.
 %further
%
%All in all,
%The contents of this section have clarified why, in both Figs.~\ref{fig:8} and ~\ref{fig:9}, solutions are possible if and only if $T>T_{off}^{min}$.
% Last but not least, it is worth observing that, for the parameter setting considered in this work, \textcolor{red}{the inequality
 %
% \begin{equation}
     %T_{off}^{min}<T_{off}^*
%     \label{ineq2}
% \end{equation}
 %
% \noindent always holds true.}
% \textcolor{black}{As a final remark, which is important in the discussion of the constrained case for $\alpha_{off(on)}\leq 2$, as will be clear shortly, it may be easily demonstrated that $T^{min}_{off(on)}$ is smaller than $T^{*}_{off(on)}$.}

  \begin{figure*}[t]
\centering
%(a)\includegraphics[width=1.75\columnwidth]{fig5a.pdf}\\ \vspace{5mm}
%(b)\includegraphics[width=1.6\columnwidth]{fig5b.pdf}
%
(a)\includegraphics[scale=.4]{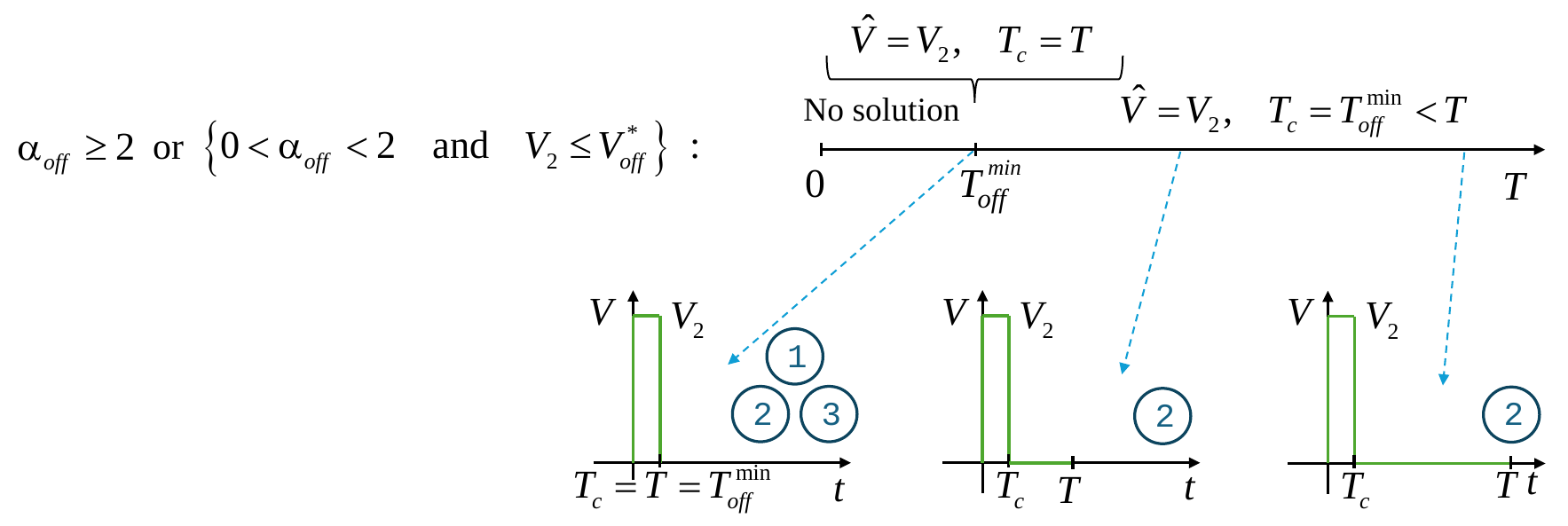}\\ \vspace{5mm}
(b)\includegraphics[scale=.4]{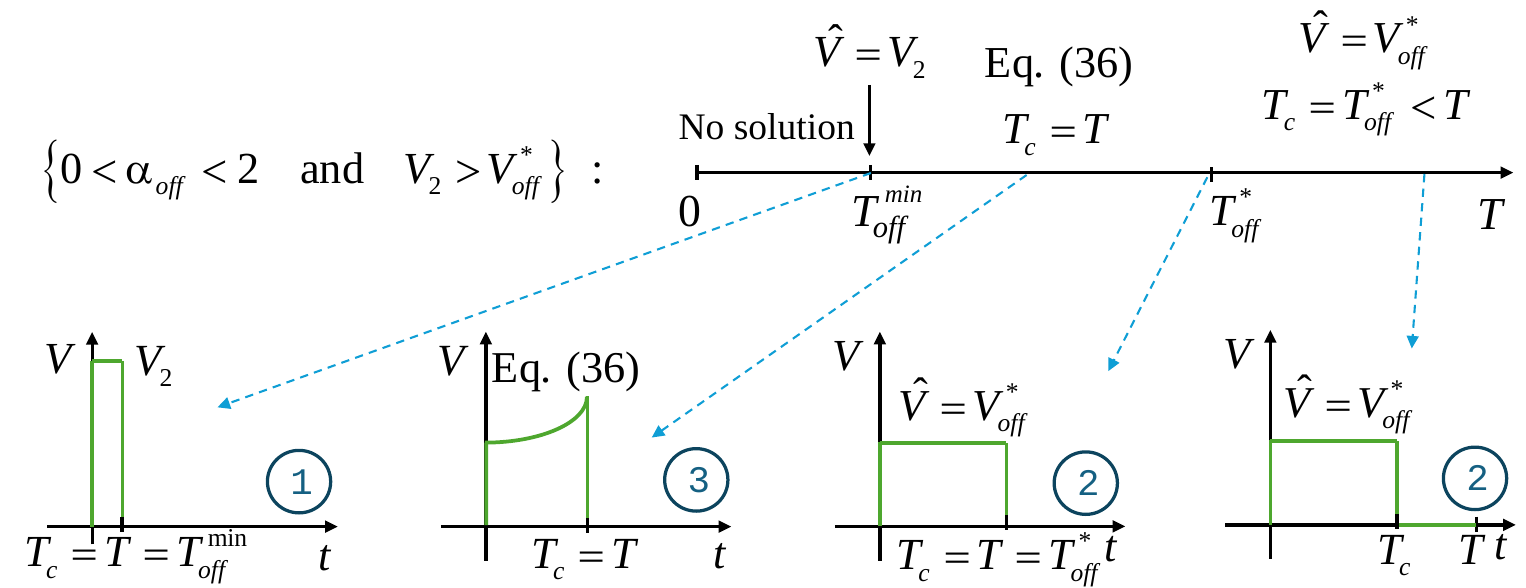}
\caption{Summary of the proposed switching control protocol for inducing the most energetically-favorable RESET
% switching
transitions across
%a memristor the VTEAM model is assumed to be fitted to.
a first-order memristive device modelled via the VTEAM mathematical description.
%More specifically,
It recommends a voltage $\hat{V(w)}$, to be applied across the device for any value the state $w$ assumes as it increases from some initial value $w_i$ to some final value $w_f$ across the respective existence domain $[w_{on},w_{off}]$, on the basis of the available programming time $T$, for each of two possible RESET control regimes, depending critically upon $\alpha_{off}$, $V_1=v_{off}$, and $V_2$.
Each of the two panels (a) and (b) includes a qualitative sketch for the time course of the optimal control voltage waveform, including a graphical indication, based on the use of numbered circles, for any case from section \ref{sec:3a3}, it refers to,
%(refer to the numbered circles),
at the critical points, separating the regions, the $T$ axis is partitioned into, and at one arbitrary point within each of these regions. In any scenario, where $T_c<T$, $\hat{V}$ is enforced to $0$ for the $(T-T_c)$-long remainder of the programming phase. A similar control scheme for reducing Joule losses across the device, the VTEAM model is fitted to, as it undergoes SET switching transitions, could be easily conceived.}
%of plots (a) and (b) of Fig. \ref{fig:7}, respectively.
%\textcolor{black}{Consult the text for the insights.}
%optimal control for the {\it off} switching of a VTEAM memristive device. A similar scheme is applicable to the {\it on} switching.
%}
\label{fig:7}
\end{figure*}

\vspace{2mm}

\section{Discussion} \label{sec:4}
% da qui 12 12 2025
In the main, we explored both the unconstrained and constrained optimization of the programming
%resistance switching
process of first-order memristive devices, the VTEAM model could be fit to. The objective was to minimize Joule losses during the resistance update procedure, a matter of invaluable practical importance nowadays.
In the unconstrained case, the optimal solution was found to depend on the value of the model parameter $\alpha_{off(on)}$. % in the VTEAM model.
When $\alpha_{off(on)}\geq 2$, the optimal solution, resulting in negligible Joule losses, %should %theoretically
consists of a voltage square pulse of %infinite
height %$V_0$
as large as possible,
%amplitude, $V_0 \rightarrow \pm \infty$, and zero duration,
%$T\rightarrow 0$.
and width as small as possible.
%, which would result in negligible Joule losses. %
%which would
%In this case, $Q_{off(on)}\rightarrow 0$.
On the contrary, for $0 < \alpha_{off(on)} < 2$, the optimal solution is a voltage square pulse of finite height %$V_{off(on)}^*=2/(2-\alpha_{off(on)})v_{off(on)}$,
$V_{off(on)}^*$,
as expressed by Eq.~(\ref{eq:13}), and finite width $T_{off(on)}^*$, as reported in Eq.
%which can be evaluated with the use of
~(\ref{eq:Toff:VTEAM:2}) ((\ref{eq:Ton:VTEAM:2})) for off (on) switching. %, respectively.
The Joule losses across the device, as it is subject to a RESET (SET) voltage stimulus of this kind, are estimated via Eq.~(\ref{eq:Qoff:VTEAM:2}) ((\ref{eq:Qon:VTEAM:2})) for off (on) switching.  %, respectively. \\
%Although these findings were derived for a square voltage pulse, we believe they have broader applicability.

Including constraints on the range $[V_1,V_2]$ of voltages, applicable across the device, and fixing the programming time $T$ %phase time interval,
adds challenges to the optimization problem. %solution.
In order to derive optimal solutions in these scenarios, we employed the principle of Pontryagin.
%In the area of constrained optimization, our findings, focusing without loss of generality on the RESET switching transitions, can be summarized as follows.
Through the examination of three cases, differing in the signs of the two Lagrange multipliers, a set of optimal solutions may be identified.
%In certain situations there exists only one suitable solution, in others, where a few solutions are admissible, the choice should fall for the most energetically-favorable one. %resulting in optimal control.
%Furthermore,
Focusing without loss of generality on the off switching scenarios, our findings, illustrated in a compact form in Fig. \ref{fig:7}, may be summarized as follows.

A successful off switching operation is impossible when the shortest RESET programming time %OFF switching time
$T_{off}^{min}$
%time interval for switching at the voltage of maximum magnitude, $V_2$,
is longer than the %programming phase time interval $T$ specified beforehand.
programming time $T$ specified beforehand.
%Firstly,
%\textcolor{black}{The solution to the %optimal control strategy
%optimization problem
%shows sensitivity
%to both model parameters and constraints,
%\textcolor{red}{
%to the constraints, defining the allowable range $[V_1,V_2]$ of voltages, applicable across the device, and the maximum admissible time interval $T$ for the programming phase, %},
%respectively.
For $T \geq T_{off}^{min}$, $\alpha_{off}$ has once again the stronger impact on the selection of the most energetically-favorable programming voltage stimulus. When its value \textcolor{black}{is larger than or equal to 2 (smaller than 2)}, the optimal solution to the Joule loss minimization problem is illustrated in Fig.~\ref{fig:8} (\ref{fig:9}).

First of all, two marginal scenarios need to be accounted for.
%situations,
%which may occur, irrespective of the value of $\alpha_{off}$, in the constrained case, need to be described.
%With reference to Figs.~\ref{fig:8}(a)-(b) and ~\ref{fig:9}(a)-(b),
\textcolor{black}{Firstly, with reference to Figs. \ref{fig:7}(a)-(b)}, irrespective of $\alpha_{off}$ and of $V_2$, when $T=T_{off}^{min}$, the only possible control voltage for resetting the device consists of a square pulse of the largest possible height $\hat{V}$, fixed to $V_2$, and width $T_c = T$.
%should be applied across the device %for the entire programming time
%to turn it off. % (refer to each of the two panels in Fig.~\ref{fig:7}).
%Also,
Secondly, with reference to plot (b) of Fig. \ref{fig:7}, when $\alpha_{off}$ is smaller than $2$,
%$V_2>V_{off}^*=2/(2-\alpha_{off})v_{off}$,
$V_2>V_{off}^*$,
and $T=T_{off}^{*}$, a square voltage pulse of height $\hat{V}$ fixed to $V_{off}^*$ and width $T_c = T$ should be applied across the device %for the entire programming time
to turn it off in the least power-consuming form.

%Secondly,
Let us proceed now with the classification of the most probable situations occurring when the inequality $T\geq T_{off}^{min}$ holds true.
%, as descends from the analysis of section\ref{sec:3a2}. %the following
\begin{enumerate}
\item If %the non-negative parameter
$\alpha_{off}\geq 2$,
%is larger than or equal to %$\geq2$
%$2$, %case 1 or 3 (case 2) from
%the study from section \ref{sec:3a2}
%optimal control strategy
it is recommended to apply a square voltage pulse of the maximum possible height $V_2$ (see Fig.~\ref{fig:8}(b)) and appropriate width $T_c = T_{off}^{min}$, smaller than the programming time $T$,
%phase time interval, %, followed by zero voltage,
across the %non-volatile
memristor (refer to Fig.~\ref{fig:7}(a)).
%, revealing how shaping the optimal control voltage this way covers also the case, where $\alpha_{off} < 2$ and $V_2 \leq V_{off}^*$).
%(see Fig. \ref{fig:7} once more).
\item  %On the other hand, assuming the condition $T > T_{off}^{min}$ to be satisfied,
\textcolor{black}{If
$\alpha_{off}<2$,}
three optimal solutions are possible. \\ %descending from the study of section \ref{sec:3a2}. %descending from the proposed optimal control strategy,
\begin{itemize}
    \item
One of them,
%descending from case 1 and 3 (case 2) of section \ref{sec:3a2},
resulting when the inequality $V_2 \leq V_{off}^*$
%$V_2 \leq V_{off}^* = 2/(2-\alpha_{off})v_{off}$
holds true, %, and illustrated once again in Fig.~\ref{fig:7}(a)),
envisages the stimulation of the device with the largest admissible voltage $V_2$
for a suitable time interval $T_c = T_{off}^{min}$ smaller than the programming time $T$ (refer once more to Fig.~\ref{fig:7}(a)).
%IMPORTANT NOTE FOR US, 12 08 2025, ALON
% Demonstration that for $\alpha_{off}< 2$ and $V_2 \leq 2/(2-\alpha_{off})v_{off}$, the optimal solutions, derived for cases 1, 2 and 3 from section \ref{sec:3a2}, agree one with the other.
%Consider the scenario, where solutions exists, i.e. when $T>T_{off}^{min}$. We have:
%\alpha_{off}<2 and V_2 \leq 2/(2-\alpha_{off})v_{off}
%-> in case 2 we choose V_2 as voltage levels cannot exceed V_2
%-> in case 3, if \tilde{V}(w)>V_2, we choose V_2 as voltage levels cannot exceed V_2, and if \tilde{V}(w)<V_2, we still choose V_2 on the basis of Fig. 2(a), where, clearly, for \alpha_{off}<2 and V_2 \leq 2/(2-\alpha_{off})v_{off}, (at V=2/(2-\alpha_{off})v_{off} each curve has a minimum), the losses for V= \tilde{V}(w) would be higher than for V=V_2
The second and third admissible solutions, illustrated in Fig.~\ref{fig:7}(b), should be considered when
$V_2 >V_{off}^*$,
%$V_2 > 2/(2-\alpha_{off})v_{off}$,
as was the case in the numerical simulation shown in Fig.~\ref{fig:9}. \\
% da qui
\item
\textcolor{black}{In regard to the second solution}, %taking into account that $T_{off}^{min}<T_{off}^*$,
occurring when $T$ is set to intermediate values between the shortest possible RESET programming time
%OFF switching time
$T_{off}^{min}$ and the time $T_{off}^*$,
at which the most energetically-favorable off switching voltage is $V_{off}^*$,
%$V_{off}^*=2/(2-\alpha_{off})v_{off}$,
%$T_{off}^{min}<T<T_{off}^{*}$,
a state-dependent
%optimal control
voltage $\hat{V}(w)$, shaped according to the formula (\ref{eq:VTEAM:Vhat:7}),
should be applied across the device for the entire programming time $T$,
%as proposed in case 3 of section \ref{sec:3a2},
providing the least power-consuming option, %\textcolor{red}{
as may be inferred from Fig. \ref{fig:2}(b) while taking into account that, here,  %satisfaction of
the inequality
%$V_{off}^{*}=2/(2-\alpha_{off})v_{off}<\min\{\tilde{V}(w),V_2\} \leq V_2$
$V_{off}^{*}<\hat{V}=\min\{\tilde{V}(w),V_2\} \leq V_2$
is satisfied for all $w$. %}
\textcolor{black}{%(see
Fig.~\ref{fig:9}(c) reveals for $\alpha_{off}=1$ how importantly does the programming time $T$
affect the dependence of the optimal voltage $\hat{V}(w)$ from Eq.~(\ref{eq:VTEAM:Vhat:7}) upon the device state $w$.
%and how,
As $T$
%Interestingly, as $T$
tends to $T_{off}^{min}$ ($T_{off}^{*}$),
%here found to be equal to $5.493$~{\textmu}s ($21.972$~{\textmu}s),
%in Fig. \ref{fig:9}(c),
$\hat{V}(w)$
%from Eq.~(\ref{eq:VTEAM:Vhat:7})
reduces to the dotted red (dotted green) solution
$V_2$ ($V_{off}^*$).
%($2/(2-\alpha_{off})v_{off}$). %).
%It is important to recall that
For any $T\in[T_{off}^{min},T_{off}^{*}]$, the value for $\lambda_1$ in the $T$-dependent formula (\ref{eq:VTEAM:Vhat:6}) for $\tilde{V}(w)$
%via Eq. %(\ref{eq:VTEAM:Vhat:7}),
%the programming time
%across the range $(T_{off}^{min},T_{off}^{*})$
%identifies
may be determined from Eq. (\ref{eq:VTEAM:cond5}), where, using Eq.~(\ref{eq:VTEAM:Vhat:7}), $\hat{V}(w)=\tilde{V}(w)$ and $T_c$ is known, being equal to $T$.
%, which coincides with $T$.
%where $\hat{V}$ is  %replaced by its expression from
%expressed via Eq. (\ref{eq:VTEAM:Vhat:7}).
%, which is identically equal to $T$.
%
%Once $\lambda_1$ is determined, Eq. (\ref{eq:VTEAM:Vhat:6}) is employed to derive $\tilde{V}$, which results in a unique $T$-dependent formula for $\hat{V}$, as reported in Eq. (\ref{eq:VTEAM:Vhat:7}).
%This, on its own, determines one particular $Tc$-long solution for $\tilde{V}(w)$, according to (\ref{eq:VTEAM:Vhat:6}), and thus for $\hat{V}(w)$, according to (\ref{eq:VTEAM:Vhat:7}).
With reference to plot (c) from Fig. \ref{fig:9}, the values for $\lambda_1$ for the first, second, third, fourth, fifth, and sixth value assigned to $T$ from the set $\{5.493,6,8,10,20,21.972\}$~{\textmu}s, in which the first (last) one in turn correspond to $T_{off}^{min}$ and $T_{off}^{*}$, are
$14\,$ mSV$^2$,
$5.652\,$ mSV$^2$,
$2.168\,$ mSV$^2$,
$1.224\,$ mSV$^2$,
$0.05589\,$ mSV$^2$,
and $0\,$mSV$^2$, respectively (note that, in the latter case, Eq. \eqref{eq:cond3} is satisfied with both factors $\lambda_1$ and $\beta[\hat{V}(w)]$ on its left hand side being identically equal to zero).
\textcolor{black}{Fig. \ref{fig:9}(d) shows the speed in the device RESET transition for values, assigned to $T$, within and above the range $(T_{off}^{min},T_{off}^*)$.
For the situation, where $T=6\;(8)$~{\textmu}s, when, at $t=2.726\;(7.234)$~{\textmu}s, the state $w$ %, shown in plot (d),
attains the value $0.629\;(0.864)$ during its
%$T_c=T=6(8)$~{\textmu}s-long transition
$T_c=T$-long transition from $w_i=0.1$ to $w_f=0.9$, the voltage stimulus,
illustrated in plot (c) and specified by Eq.~(\ref{eq:VTEAM:Vhat:7}), transitions from $\tilde{V}(w)$ to the fixed value $V_2=5$~V.\\} % , here equal to $5$V.
%according to the parameter setting specified in the caption of Fig. \ref{fig:8}.}
}
%, %expressed via %by $\tilde{V}(w)$ from
%Eq.~(\ref{eq:VTEAM:Vhat:7}), %in case 3,
%against the device internal state $w$
%in scenarios where the programming time
%phase time interval
%$T$ is set to intermediate values between the shortest possible OFF switching time $T_{off}^{min}$ and the time $T_{off}^*$. %, according to case 2.
%
%Importantly,
% Note: \tilde{V}(w) is larger than 2/(2-\alpha_{off})v_{off}, but may also go larger than V_2, this is why \hat{V} is clipped to V_2 in some cases
% (refer to Fig. 4(c) for T=6\mus or also for T=8\mus
\item
\textcolor{black}{In regard to the third solution}, occurring when $T>T_{off}^*$, the application of the constant voltage
$\hat{V}=V_{off}^*$
% $\hat{V}=2/(2-\alpha_{off})v_{off}$
(see Fig.~\ref{fig:9}(b)) across the device over a $T_c=T_{off}^*$-long part of the programming phase,
%as resulting from case 2 of section \ref{sec:3a2},
is the most energetically-favorable option, as inferable from Fig. \ref{fig:2}(b).
\end{itemize}
\end{enumerate}
%with a state-dependent voltage waveform, according to equation (\ref{eq:VTEAM:Vhat:7}), throughout the programming phase, when $T_{off}^{min} <T<T_{off}^*$, or with an intermediate voltage level, specifically $2/(2-\alpha_{off})$, for a suitable time interval $T_c \leq T$, when $T>T_{off}^*$ (refer to Fig.~\ref{fig:7}(b)).
%
%\textcolor{red}{Here, as anticipated in the caption of Fig. \ref{fig:7}, we should explain why the solution $\hat{V}(w)=\min\{V_2,\tilde{V}(w)\}$ from case 3 is not an optimal choice for $\alpha_{off}<2$ in panel (a).}
%
%(one is shown in Fig.~\ref{fig:7}(a), and two -- in Fig.~\ref{fig:7}(b)).
%For a complete picture, see Fig.~\ref{fig:7}.
%$T\geq T_{off}^{min}$

\textcolor{black}{
Importantly, in the scenarios, panel (a) ((b)) of Fig. \ref{fig:7} corresponds to,
the RESET switching energy keeps equal to its smallest possible value (decreases monotonically from its largest possible value to its smallest possible value) as the programming time $T$ is progressively increased from the shortest RESET programming time
%OFF switching time
$T_{off}^{min}$.
Our optimization strategy enables considerable switching energy savings over traditional programming methods, as illustrated in Figs.~\ref{fig:8}(a) and \ref{fig:9}(a), which compare the Joule losses across the VTEAM device, under constant voltage-based and optimal voltage-based switching control protocols for a couple of scenarios from the case, where $\alpha_{off}\geq 2$, and for one scenario from the case, where $\alpha_{off}< 2$, respectively.
As shown in Fig.~\ref{fig:8}(a), employing the optimal switching control method for $\alpha_{off} =3$ result in RESET %switching
transition energy savings beyond one order of magnitude for sufficiently long programming times.
Referring now to Fig.~\ref{fig:9}(a), the RESET transition
%switching
energy savings, resulting from the application of the proposed optimal switching control strategy for $\alpha_{off}=1$ appear to be modest within the interval $T_{off}^{min} \leq T \leq T^*_{off}$, but rise significantly when $T > T^*_{off}$. }

The proposed theoretical approach to minimize Joule losses in the programming phase of first-order memristors has wide applicability (see the Appendix to learn how it shapes for a different memristive model~\cite{Miranda20a}).

%Last but not least, as a proof of evidence for the generality of the theoretical approach, introduced in this paper, the Appendix includes another %practical
%application of the proposed energy loss minimization methodology, %in the Appendix,
%illustrating the benefits of our optimal switching control protocol for a first-order memristive device \textcolor{black}{with weak volatility,}
%with good data retention capability,
%described through the dynamic balance model~\cite{Miranda20a} from Miranda and Su$\tilde{\textnormal{n}}$\'e.
%Despite certain similarities exists between the switching control protocols,
%synthesised for the VTEAM and dynamic balance models, as may be evinced by comparing the respective guidelines, illustrated in Figs. ~\ref{fig:7} and \ref{fig:A3}, respectively,
%the application of the method of Lagrangian multipliers is model- and constraint-dependent.
%some of the guidelines, it specifies, may be found for
%Therefore, it is not possible to identify a unique recipe,
%single switching control recipe,
%establishing the properties, a programming stimulus should feature,
%for the Joule losses, it would induce across a memristive device of any kind, to be minimized.
%under general design specifications.

%In principle, this derivation demonstrates that our method can be relatively easily applied to devices described by other first-order memristive models. Furthermore, we emphasize that the optimal switching protocols derived in this work are model- and constraint-dependent, although certain similarities can be seen comparing Figs.~\ref{fig:7} and \ref{fig:A3}.

\section{Conclusions}\label{sec:5}
In this paper, we have presented a rigorous theoretical approach to reduce the switching energy for first-order non-volatile %ReRAM
memristors % devices
%, characterized by highly-nonlinear operating principles \cite{asc_adv_ele_mat_2022},
to the bare minimum.
%In order to support the theoretical analysis with illustrative examples,
%the proposed methodology has been specifically applied to memristors, admitting two mathematical descriptions, specifically the \textcolor{black}{Voltage ThrEshold Adaptive Memristor} model in the main text and the dynamic balance model in the Appendix, resulting in the
%synthesis of optimal switching control measures for each of the two case studies.
%where we have pinpointed the optimal control strategies.
The control signals, shaped through our optimization strategy for inducing the least power-hungry resistance switching transitions across a device, depend crucially upon the underlying switching kinetics as well as upon the design constraints on the programming time and on the minimum and maximum allowable stimuli.
The adoption of the proposed Joule loss minimization methodology may result in considerable energy savings during the programming phase of large arrays of memristors employed for in-memory-computing applications nowadays.
The findings disclosed in this manuscript open up a variety of intriguing opportunities for future research.
%
%The methodology could be applied to volatile memristive devices with non-monotonic DC current versus voltage characteristics for minimizing the energy to be supplied continuously for polarizing them on some negative differential resistance bias point.
Seeking the experimental validation~\cite{fleck2016energy} of the theoretical predictions, is %definitely
one of the priorities in our research agenda.
%On one side, expanding our results to encompass other memristive models holds promise.
%
%On another front, it is important to experimentally validate our theoretical predictions.
%For this purpose, observing the actual behavior of a device as per the VTEAM model (or any other well-defined first-order memristive model) is crucial.
%
As additional promising avenue for future research, measuring Joule losses across a memristive device, during SET and RESET transitions, may provide valuable insights for the development of an opportune model, predicting its nonlinear dynamics, and for the estimation of its parameters.
On a different scientific front,
an optimization technique of the kind described in this paper
%a similar
%optimization technique
could be leveraged %\emph{mutatis mutandis} %applied to
to estimate the minimum energy costs in  innovative memristor programming
%approaches to
%associated to novel approaches to memristor  programming
procedures, such as the methodology, originally introduced in \cite{Pershin_2019}, and further explored in \cite{pershin_2019_bif_anal_TaO_mod}, and \cite{messaris2023},
where periodic signals at high frequency were employed to induce controlled resistance switching in a non-volatile memristor device endowed with fading memory \cite{ascoli2016}.
%\textcolor{black}{It is important to observe, as a very final remark, that the optimization strategy, proposed in this paper, is of general applicability, and may potentially allow considerable energy savings in the programming phase of large arrays of memristors employed for in-memory-computing applications nowadays.}
\textcolor{black}{From a mathematical perspective, identifying the optimal switching strategies for fractional-order memristive models~\cite{astin2025low} would be interesting.}

\section*{Acknowledgement}
The authors gratefully acknowledge J.~P.~Strachan and S.~Menzel
for their helpful discussion. YVP was supported by the NSF grant EFRI-2318139.

%\appendix %\label{app:example}
\section*{Appendix}

\renewcommand\thefigure{A\arabic{figure}}
\setcounter{figure}{0}

\renewcommand\theequation{A\arabic{equation}}
\setcounter{equation}{0}

The mathematical description, the proposed memristor switching energy minimization procedure is applied to in this Appendix, is the so-called
dynamic balance model~\cite{Miranda20a}. Proposed by Miranda and Su$\tilde{\textnormal{n}}$\'e in 2020,  \textcolor{black}{it falls in the class of first-order voltage-controlled volatile memristors.
If the duration of the programming operation is much shorter than the intrinsic time constants of the model, the system can be treated using a non-volatile approximation. This is the situation examined in this work.}

\subsection{Dynamic balance model}\label{sec:A1}
The dynamic balance model
employs the following \textcolor{black}{continuous and differentiable}
ODE to govern
%Within the dynamic balance model,
the time evolution of the memristor internal state variable $x$:
%is described by the first-order ordinary differential equation
\begin{equation}
    \frac{\textnormal{d}x}{\textnormal{d}t}=f(x,V)=\frac{1-x}{\tau_S(V)}-\frac{x}{\tau_R(V)}.
    \label{eq:Mm2}
\end{equation}
%In Eq.~(\ref{eq:Mm2}),
Here $x$ is confined to the interval $[0,1]$, whereas $\tau_{S(R)}(V)$ stands for the SET (RESET) time constant, which depends exponentially upon the voltage $V$ let fall across the device~\cite{Miranda20a}.
In this work, we employ the very same expression for $\tau_{S(R)}(V)$, as reported in~\cite{Miranda20a}, while ensuring its parameters, specifically the intrinsic (zero voltage) time constant $\tau_{0,S(R)}$ as well as the constant $\eta_{S(R)}$, %it contains,
to admit appropriate physical units.
The voltage-dependent SET (RESET) time constant $\tau_{S(R)}(V)$ is thus described via

%dimensionality:
\begin{equation}
    \tau_{S(R)}(V)=\tau_{0,S(R)}e^{-\eta_{S(R)}V}, \label{eq:taus}
\end{equation}

\noindent where $\tau_{0,S(R)}>0$,
%($\eta_R<0$), while
whereas %the SET (RESET) scaling factor
$\eta_{S(R)}>(<)0$.
The current through the memristor is defined through the state-dependent Ohm's law

\begin{equation}
I=G_M(x)V=\left[ (1-x)G_{\textnormal{min}}+xG_{\textnormal{max}}\right]V,
\label{eq:Mm1}
\end{equation}
\noindent where $G_{min}$ and $G_{max}$ are the minimum and maximum allowable memductance values, respectively.
%
% da qui 1234

\textcolor{black}{Importantly, in the analysis to follow, it is assumed that $\tau_{0,S}(\tau_{0,R})\gg T$. Under this assumption, in our optimization problem, it is reasonable to neglect the first (second) exponentially-small SET (RESET) addend in the formula for $f(x,V)$ in Eq.~(\ref{eq:Mm2}) for some finite $V<(>)0$ and both addends when $V=0$.}

\textcolor{black}{Disregarding the first (second) SET (RESET) additive term on the right-hand side of Eq. (\ref{eq:Mm2}), while analysing the device switching kinetics under a voltage pulse,
the memory state $x$ is found to undergo a monotonic decrease (increase) toward its lower (upper) bound $0$ ($1$), while its memductance is concurrently subject to a progressive reduction (growth) toward its minimum (maximum) value $G_{min}$ ($G_{max})$, which is essentially what happens during the off (on) resistance switching transition in a non-volatile device.
Additionally, %with such a hypothesis,
the state evolution function of the approximate dynamic balance model features unique and complementary signs, irrespective of the state, for non-zero voltages of opposite signs.%, as was the case for the VTEAM model when the modulus of the voltage was chosen larger than the maximum between the positive and negative threshold voltages.
}
\vspace{-.375cm}
\subsection{Optimal Unconstrained Solution} \label{sec:A2}
%\label{sec:3b1}
We initially estimate the % switching
energy costs to be paid for programming a memristor, the dynamic balance model DAE set \eqref{eq:Mm2}-\eqref{eq:Mm1} is fitted to, while following a constant voltage-based switching control protocol, whereby no constraint is enforced on
%which is synthesized below,
%without imposing
%optimal switching with the use of a square voltage pulse without imposing any
%
%limitations on
the amplitude and duration of the pulse let fall across the device.
Let us first consider a SET transition, induced across the device to increase its state $x$ from $x_i$ to $x_f$, by letting  a constant positive voltage $V_0$ fall across it for a finite time.
\textcolor{black}{Exploiting the analytical tractability of the Miranda and Su$\tilde{\textnormal{n}}$\'e model,} and
neglecting the second term in the sum on the right-hand side of the equation of motion~(\ref{eq:Mm2}), the first, second, and third formulas in the triplet to follow express the dependence of the internal state $x$, \textcolor{black}{renamed here $x_S$}, upon time $t$, of the pulse duration $T_S$ upon pulse height $V_0$, and of the SET switching energy $Q_S$ upon pulse height $V_0$, respectively.
\begin{eqnarray}
    x_S(t)&=&1+(x_i-1)e^{-\frac{t\textcolor{black}{-t_i}}{\tau_S(V_0)}} \label{eq:MS1}\; \\
    T_{S}(V_0)&=&\tau_S(V_0)\ln\frac{1-x_i}{1-x_f} \; , \label{eq:MS2}\\
    Q_{S}(V_0)&=&\tau_S(V_0) V_0^2 \bigg( G_{max}\ln\frac{1-x_i}{1-x_f}-  \nonumber \\
    && (G_{max}-G_{min}) (x_f-x_i) \bigg). \;\;\;\;\;\; \label{eq:MS3}
\end{eqnarray}
Similarly, assuming now to apply a negative RESET voltage $V_0$ between the two terminals of the device for determining a decrease in its state $x$ from $x_i$ to $x_f$, the predictions of the dynamic balance model, drawn while neglecting the first term in the sum on the right-hand side of the equation of motion~(\ref{eq:Mm2}), for the dependence of the state $x$, \textcolor{black}{renamed here $x_R$}, upon time $t$, of the pulse duration $T_R$ upon pulse height $V_0$, and of the RESET switching energy $Q_R$ upon pulse height $V_0$ may be evinced from the first, second, and third closed-form expressions to follow, respectively:

\begin{eqnarray}
    x_R(t)&=&x_ie^{-\frac{t\textcolor{black}{-t_i}}{\tau_R(V_0)}}  \; , \label{eq:MS1a} \\
    T_{R}(V_0)&=&\tau_R(V_0)\ln\frac{x_i}{x_f} \; , \label{eq:MS2a} \\
   Q_R(V_0)&=&\tau_R(V_0) V_0^2 \bigg( G_{min}\ln\frac{x_i}{x_f}- \nonumber \\
    && \;\; (G_{max}-G_{min}) (x_f-x_i) \bigg) \;. \;\; \label{eq:MS3a}
\end{eqnarray}

\begin{figure}[t]
\centering
\includegraphics[scale=.3]{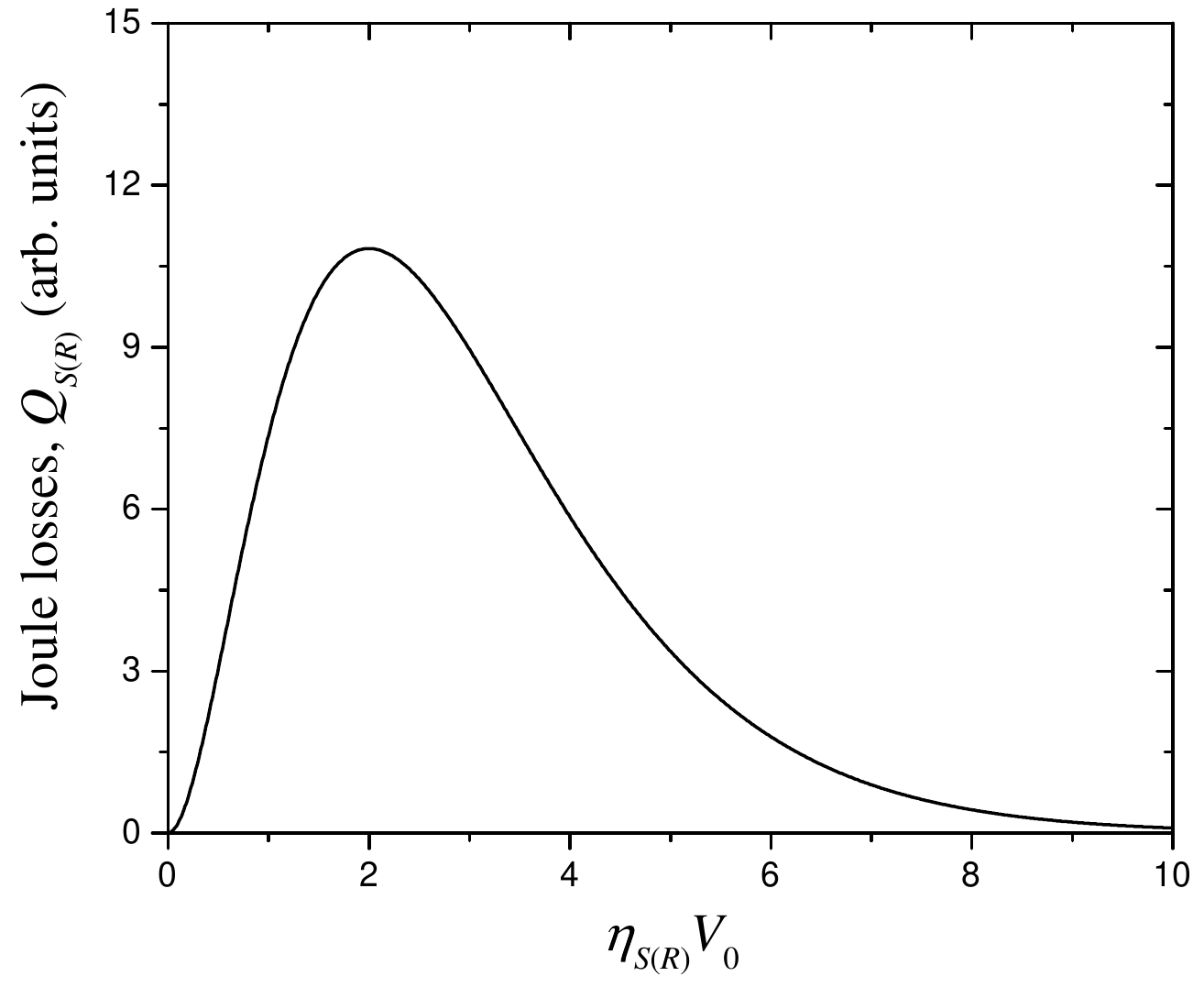}
\caption{SET (RESET) switching energy $Q_{S(R)}$%, scaled down by the factor $\tau_S(V_0) \cdot  \left( G_{min}\ln(x_i/x_f)-(G_{max}-G_{min})(x_f-x_i)\right)$,
as a function of the %\textcolor{red}{
modulus of the positive (negative) pulse height $V_0$ %}
for a first-order memristive device, characterized through the %first-order
dynamic balance model, as resulting from the closed-form expression in Eq.~(\ref{eq:MS3}) ((\ref{eq:MS3a})) for $x_i=0.1$ and $x_f=0.9$.}
%This plot was obtained using Eq.~(\ref{eq:MS3}).}
\label{fig:A1}
\end{figure}

% da qui 15 12 2025
With reference to Eq.~(\ref{eq:MS3}), the function $Q_S(V_0)$ is equal to $0$ (approaches $0$ asymptotically) for $V_0=0$ (as $V_0$ tends to positive infinity), with a maximum at $V_0=2/\eta_S$ (see Fig.\ref{fig:A1}).
Discarding the solution at/near $V_0=0$, %, which would require an infinitely-long programming time,
the SET switching energy $Q_S$ is minimized as $V_0$ is let tend to positive infinity, which would reduce the SET switching time $T_S$ from Eq.~\eqref{eq:MS2} to $0$.
Similarly, through the analysis of Eq.~(\ref{eq:MS3a}), it is easy to conclude that the RESET switching energy may be minimized as $V_0$ is let tend to negative infinity,  which would reduce the RESET switching time $T_R$ from Eq.~\eqref{eq:MS2a} to $0$.

%$V_0\in[0,\infty\}$ has a maximum at $V_0=2/\eta_S$, and is equal to zero at its endpoints (assymptotically); see Fig.~\ref{fig:A1}. Rejecting the long time solution ($V_0=0$), $Q_S$ is minimized in the limit of $V_0\rightarrow \infty$. Similarly, $Q_R$ is minimized in the limit of $V_0\rightarrow -\infty$. In these limits, $Q^*_{S(R)}\rightarrow 0$ and $T^*_{S(R)}\rightarrow 0$.

\subsection{Optimal Constrained Solution}\label{sec:A3}

Here, referring to Pontryagin's principle, the most energetically-favorable solution for the stimulus is derived by solving the set of three constraints, reported in Eqs.~(\ref{eq:cond2}), (\ref{eq:Argmin:1}), and (\ref{eq:cond3}), while concurrently taking into account design-based limitations on programming time and voltage levels.
%}
%For the {\it set} (S) transition that we consider here in detail, the Lagrangian is written as
Without loss of generality, the analysis is focused on SET switching transitions, when $V>0$, while disregarding the second RESET term on the right-hand side of the equation of motion
(\ref{eq:Mm2}).
%state evolution function $f(x,V)$ was approximated by neglecting the second RESET term on the right-hand side of the equation of motion (\ref{eq:Mm2})
%into consideration,
Similar considerations may be drawn \emph{mutatis mutandis} on RESET switching transitions, when $V<0$, while disregarding the first SET term on the right-hand side of the equation of motion (\ref{eq:Mm2}).

\begin{figure*}%[tb]
[h!]
\centering
%(a)\includegraphics[width=1.75\columnwidth]{figA3a.pdf}\\
%(b)\includegraphics[width=1.75\columnwidth]{figA3b.pdf}
%
(a)\includegraphics[scale=.37]{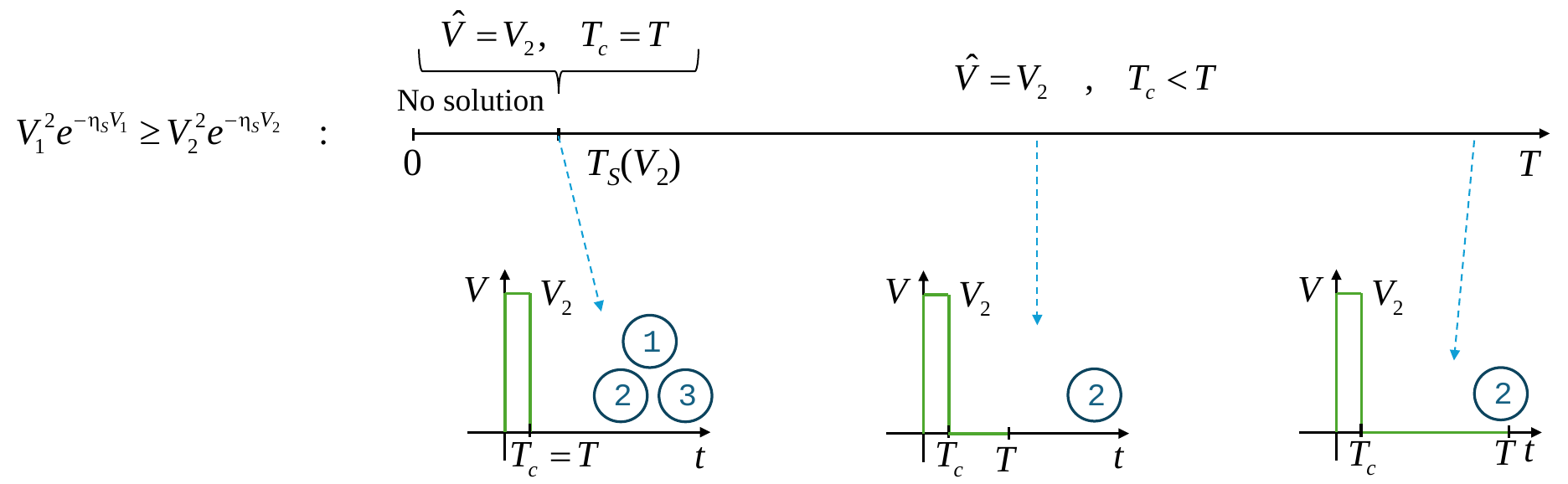}\\
(b)\includegraphics[scale=.37]{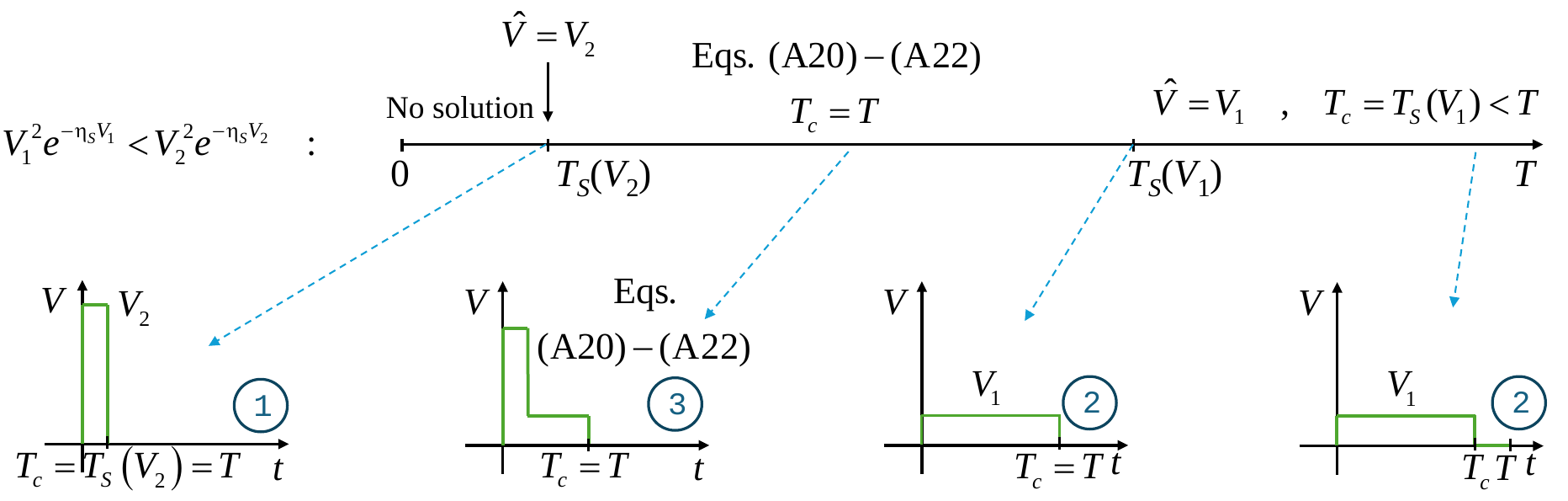}
\caption{
Classification of the % most suitable
voltage signals, which, according to the proposed control strategy, should be applied across a first-order memristor, the dynamic balance model is fitted to, for inducing the most energetically-favorable increase in its internal state from some initial value $x_i$ to some final value $x_f$, as the programming time $T$, specified beforehand, is progressively increased from the shortest SET programming time
% ON switching time
$T_s^{min} = T_s(V_2)$, under the assumption that the choice of the minimum and maximum voltage levels, $V_1$ and $V_2$ respectively, endows the exponential polynomial %$p(V_1,V_2)$ from equation (\ref{poly_for_Fig_A3})
$V_1^2 e^{-\eta_SV_1} - V_2^2 e^{-\eta_SV_2}$ \textcolor{black}{with (a) a non-negative value or (b) a negative value.
The optimal control voltage waveform $\hat{V}$ over time is qualitatively sketched,
at selected values of $T$.
}
}
%
%Based on the values of $V_1$ and $V_2$, representing the minimum and maximum control voltages, either (a) or (b) is chosen.
%Recall that the limit $T \ll \tau_{0,S}$ was assumed during the derivation. Moreover, the function $T_S(V)$ is defined by Eq.~(\ref{eq:MS2}). The same methodology applies to {\it off}-switching.}
\label{fig:A3}
\end{figure*}

\textcolor{black}{Let us begin the analysis by noting that, with $V$ confined to the range $[V_1,V_2]$,
in case $V_1^2 e^{-\eta_S V_1}-V_2^2 e^{-\eta_S V_2}$ is non-negative (negative), the locus of the SET switching energy $Q_S$ versus the pulse height $V_0$ from Eq. (\ref{eq:MS3})
%for $V_0=V$,
will admit its smallest possible value when $V_0$ is set to its upper (lower) bound $V_2$ ($V_1$) (see also plot (a) ((b)) in Fig. \ref{fig:A3}). Looking at the $Q_S$ versus $V_0$ locus in the inset of Fig. \ref{fig:A4}, drawn for $\eta=5$~V$^{-1}$, the second of these two cases would appear in case $V_1$ and $V_2$ were in turn set to $0.3$~V and $0.5$~V.\\}
Adapting Eq.~(\ref{eq:Lagrangian:1}) to the dynamic balance model, the Lagrangian $L(x,V)$ is found to admit the form

\small{
\begin{equation}
    L(x,V)=\lambda_0\frac{G_M(x)V^2(x)\tau_S(V)}{1-x}+\lambda_1\left[ \frac{\tau_S(V)}{1-x}-\frac{T}{x_f-x_i}\right]. %,
    \label{eq:Lagrangian:2}
\end{equation}
}
\normalsize{}

%\noindent where the .
Three are the possible cases that may occur, %to be analysed separately,
depending upon the choice for the Lagrange multiplier pair $(\lambda_0,\lambda_1)$. Let us analyse them separately. \\

{\bf Case 1: $\lambda_0=0$, $\lambda_1>0$}.
%Inserting the first term in the sum on the right-hand side of the state equation (\ref{eq:Mm2}) in place for $f(x,V)$ within the formula for the Lagrangian, reported in (\eqref{eq:Lagrangian:2}),
Discarding the constant term in the formula (\ref{eq:Lagrangian:2}) for the Lagrangian, the optimal positive SET voltage $\hat{V}(x)$ is computed via Eq.~(\ref{eq:Argmin:1}), which takes the form
%and disregarding the constant term in the above Lagrangian, we obtain

\begin{equation}
    \hat{V}(x)=\underset{V_{1}\leq V\leq V_2}{\textnormal{Argmin}}  \frac{\tau_S(V)\lambda_1}{ 1-x}. %=V_2.
    \label{eq:Mm:Vhat:1}
\end{equation}

Clearly, %the optimal control voltage
$\hat{V}(x)$ should be set to the upper bound $V_2$ in the device voltage admissible range $[V_1,V_2]$, whose lower bound $V_1$ should be carefully selected, as discussed shortly.
Solving the state %time
integral of the reciprocal of $f(x,V)$ from Eq.~(\ref{eq:5a}), the formula for the \textcolor{black}{effective} switching time is found to read as

\begin{equation}
   T_c=\tau_S(V_2)\ln \frac{1-x_i}{1-x_f}, %\equiv T,
    \label{eq:Mm:cond1}
\end{equation}

\noindent which is equivalent to the programming time $T$, as established by the constraint~(\ref{eq:cond3}).
All in all, in case 1, the highest possible voltage %$\hat{V}=V_2$ is
should be applied across the memristor during the entire programming phase, which should stretch over a temporal window of width computable via Eq.~(\ref{eq:Mm:cond1}),
which %, rather interestingly,
coincides with $T_S(V_0)$, computable via Eq.~(\ref{eq:MS2}) for $V_0=V_2$, denoting the shortest SET programming time
%ON switching time
$T_S^{min}$ for the memristor device characterized by the dynamic balance model. \\
%As this case may only occur for a single programming time, its occurrence is improbable. \\

%time interval $t_c=T$ whose value is given by equation~(\ref{eq:Mm:cond1}). We note that equation~(\ref{eq:Mm:cond1}) agrees with Eq.~(\ref{eq:MS2}).
%da qui

{\bf Case 2: $\lambda_0>0$, $\lambda_1=0$}.
Here, setting arbitrarily $\lambda_0$ to $1$ in the formula (\ref{eq:Lagrangian:2}) for the Lagrangian, Eq.~(\ref{eq:Argmin:1}), whose solution provides the optimal positive SET voltage $\hat{V}(x)$, assumes the form

%\begin{equation}
%    \hat{V}(x)=\underset{0\leq V\leq V_2}{\textnormal{Argmin}}  \frac{\tau_S(V)G_M(x)V^2}{ (1-x)}=0.
%    \label{eq:Mm:Vhat:2}
%\end{equation}

\begin{eqnarray}
    \hat{V}(x)&=&\underset{V_1\leq V\leq V_2}{\textnormal{Argmin}}  \frac{\tau_S(V)G_M(x)V^2}{ (1-x)}. %=
    %\nonumber
    %\\
     \label{eq:Mm:Vhat:2}
\end{eqnarray}

Solving Eq.~(\ref{eq:Mm:Vhat:2}), the most energetically-favorable control voltage $\hat{V}$ should be selected via

\begin{eqnarray}
       \hat{V}&=&  \begin{cases}
      V_1, & \textnormal{if}\quad V_1^2e^{-\eta_SV_1} %\leq
      \textcolor{black}{<}V_2^2e^{-\eta_SV_2},\\
      V_2, & \textnormal{if}\quad V_1^2e^{-\eta_SV_1}\geq V_2^2e^{-\eta_SV_2}.
    \end{cases}
    \label{eq:Mm:Vhat:2a}
\end{eqnarray}

%Since $\lambda_1=0$,
%On the basis of condition (\ref{eq:cond3}),
Solving the %time
state integral of the reciprocal of $f(x,V)$ in Eq.~(\ref{eq:5a}), the \textcolor{black}{effective} switching time $T_c$, expected to be smaller than or equal to the
programming time $T$ in view of condition (\ref{eq:cond3}),
may be set to one of two admissible values on the basis of an inequality condition, as reported here:

\small{
\begin{eqnarray}
       T_c&=&  \begin{cases}
      T_S(V_1), & \textnormal{if}\quad V_1^2e^{-\eta_SV_1} %\leq
      \textcolor{black}{<} V_2^2e^{-\eta_SV_2},\\
      T_S(V_2), & \textnormal{if}\quad V_1^2e^{-\eta_SV_1}\geq V_2^2e^{-\eta_SV_2}, %.
    \end{cases}
    \label{eq:Mm:T_C:2a}
\end{eqnarray}
  }
\normalsize{}

\noindent \textcolor{black}{where $T_S(V_1)>T_S(V_2)$.}
%\noindent Irrespective of its value, it would be smaller than .
All in all, in case 2 the optimal switching control voltage should be selected according to Eq. (\ref{eq:Mm:Vhat:2a}) across an initial fraction of the programming time interval, or throughout it.
%or over a fraction of the associated $T$-long time interval.
\textcolor{black}{In the first case, the control voltage should be forced to zero
%$T_c$-long part
in the $(T-T_c)$-long remainder of the programming phase.} \\

%over the remainder of the $T$-long time interval,
%or throughout the .
%
%
%Consequently, the timing requirement here is that $T_c \leq T$.

%However, the switching time must be such that $t_c = T$. In fact, the solution $\hat{V}=0$ lacks physical significance, because at zero applied voltage, both components in the equation of motion (\ref{eq:Mm2}) become relevant. In this scenario, the memristor will settle in an intermediate state, such as $x=0.5$, assuming symmetric constants.

{\bf Case 3: $\lambda_0>0$, $\lambda_1>0$}.
Keeping a unitary value for $\lambda_0$ in the Lagrangian, expressed via (\ref{eq:Lagrangian:2}), Eq.~(\ref{eq:Argmin:1}) takes here the form

\begin{equation}
        \hat{V}(x)=\underset{V_1 \leq V\leq V_2}{\textnormal{Argmin}}  g(x,V), \label{eq:Mm:Vhat:3}
\end{equation}

\noindent where $g(x,V)$ is defined as

\begin{equation}
g(x,V) = \frac{\tau_S(V)(G_M(x)V^2+\lambda_1)}{ 1-x}.
\label{def_g}
\end{equation}

If it exists, the optimal positive SET voltage, minimizing the argument of the Argmin operator in Eq.~ (\ref{eq:Mm:Vhat:3}), is a root of the second-order polynomial

\begin{equation}
\eta_SG_M(x)V^2(x)-2G_M(x)V(x)+\lambda_1 \eta_S=0, \label{2nd_order_poly_miranda}
\end{equation}

\noindent whose solutions are

\begin{equation}
    V_\pm(x)=\frac{1}{\eta_S}\pm\sqrt{\frac{1}{\eta_S^2}-\frac{\lambda_1}{G_M(x)}}. \label{eq:Mm:Vhat:4}
\end{equation}

%\begin{figure}[tb]
%\centering
%\includegraphics[width=0.95\columnwidth]{fig6.pdf}
%\caption{An example of $\hat{V}(t)$  and $G(x)$. This plot was obtained for the case of the dynamic balance model. \hl{Can be improved. Parameters?}}
%\label{fig:6}
%\end{figure}

For any given $x\in[0,1]$, when $0<G_M(x)/\eta_S^2 < \textcolor{black}{(>)}\lambda_1$,
%(\geq) \lambda_1$,
%\textcolor{red}{for any $x\in[0,1]$},
these two solutions are complex (real).
%When $G_M((x))/\eta_S^2>\lambda_1>0$ for any $x\in[0,1]$, these two solutions are real. %It can be demonstrated that
In the latter case, the specific real-valued solution, obtained by %selecting
choosing the negative (positive) sign in Eq.~ (\ref{eq:Mm:Vhat:4}), corresponds to a local  minimum (maximum) for $g(x,V)$. %whereas the other solution represents a minimum.

\begin{figure}[tb]
\centering
\includegraphics[scale=.3]{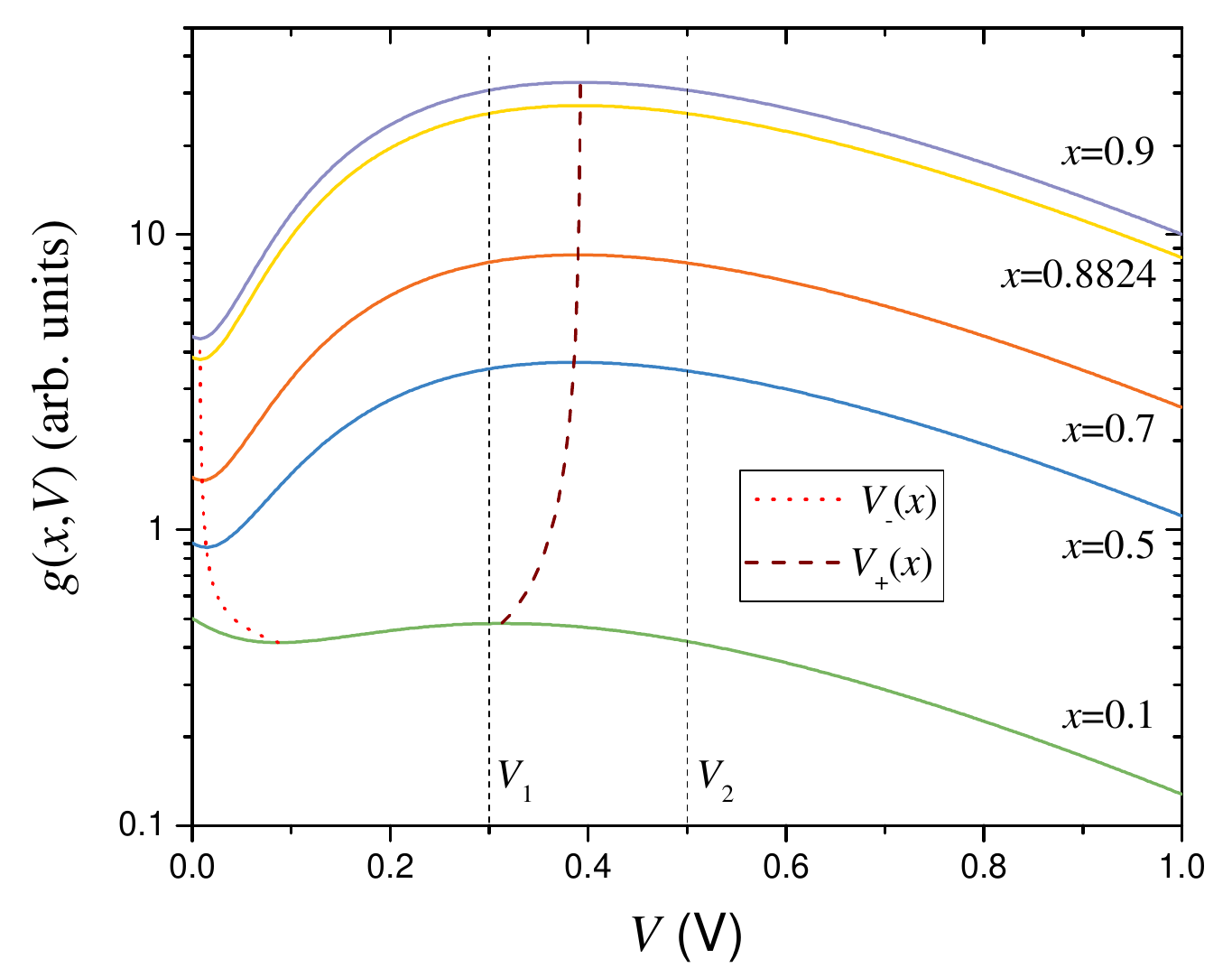}
\caption{
%Schematics of the function in Eq.~(\ref{eq:Mm:Vhat:3})
Coloured solid traces: Loci illustrating %showing %qualitatively
the dependence of the function $g(x,V)$, defined in Eq.~(\ref{def_g}), upon the voltage $V$, falling across the device, for a number of values, specifically
%0.2, 0.3, 0.4, 0.792, and 0.9,
$0.1$, $0.5$, $0.7$, $0.8824$, $0.9$,
%a given value,
%assumed progressively by the
assigned to its internal state $x$. In fact these are values $x$ progressively assumes as it increases monotonically from $x_i=0.1$ to $x_f=0.9$
%-- see the blue solid trace in Fig. \ref{fig:A5} -- %
%over the respective existence domain $[0,1]$
%in a numerical simulation, Fig. \ref{fig:A5} refers to,
\textcolor{black}{in Fig. \ref{fig:A5}, recorded in a
%retrieved from a
numerical simulation,
%to be discussed in section \emph{A.D},
%showing an %exemplary
where our optimal SET switching control strategy was applied in a scenario from case 3, where, assuming $V_1^2 e^{-\eta_S V_1}-V_2^2 e^{-\eta_S V_2}<0$ and $T\in[T_S(V_2),T_S(V_1)]$ (refer to Fig.~\ref{fig:A3}(b)),
%for a scenario, which, assuming $V_1^2 e^{-\eta_S V_1}-V_2^2 e^{-\eta_S V_2}<0$, falls in case 3, and where, upon choosing the programming time $T$ inside the range $(T_S(V_2),T_S(V_1))$,
the least power-consuming voltage stimulus $\hat{V}$ was found to transition only once, from $V_2$ to $V_1$, over the effective switching time interval.}
%, from case 3 for $V_1=0.3$~V and $V_2=0.5$~V.
%in a certain regime of parameters.
Red dotted (brown dashed) trace: Locus of the local minimum (local maximum) $V_{-(+)}(x)$ of $g(x,V)$ from Eq.~(\ref{eq:Mm:Vhat:4}) for state values in the range $[0.1,0.9]$, employing values for $\eta_S$, $\lambda_1$, $G_{min}$ and $G_{max}$ as reported in the caption of
%in the simulation %.}
%in the simulation
Fig. \ref{fig:A5}.} % refers to.}
\label{fig:A2}
\end{figure}

Consequently, %various scenarios must be thoroughly examined.
for any given state value $x\in[0,1]$, a number of distinct scenarios, leading to different optimal solutions, may occur, as listed below: %

\begin{itemize}
    \item When $G_M(x)/\eta_S^2 \textcolor{black}{<}\lambda_1$,
   % \leq \lambda_1$,
    %the function in Eq.~($\ref{eq:Mm:Vhat:3}$), let us call it $g(V)$,
    the function $g(x,V)$ from (\ref{def_g})
    is a monotonically-decreasing function of $V$,
    %(see, for example, the trace of $g(x,V)$, associated to the label $x=0.2$, in Fig. \ref{fig:A2}, to be discussed in detail later on),
    %(as may be evinced via eye inspection, this happens in Fig.~\ref{fig:A2}, to be discussed in detail later on, %inspecting , it can be realized that
    %for any state value within the set $(0,0.3)$),
    and, %(at a given $x$).
    as a result, it assumes the lowest value at the right end of the allowable control voltage interval $[V_1,V_2]$. In this scenario, the optimal switching control voltage $\hat{V}(x)$, %independent of the state $x$,
    for the given state $x$
    should % then
    be specified according to

    \begin{equation}
        \hat{V}(x)=V_2.
        \label{eq:Mm:Vhat:5}
    \end{equation}
    \item If the condition $G_M(x)/\eta_S^2>\lambda_1$ holds true,
    \textcolor{black}{as is the case in Fig.~\ref{fig:A2},
    %to be discussed in more detail in section \emph{A.D}, and
    showing the locus of $g(x,V)$
    versus $V$ for a number of values assigned to the internal state $x$ within the set $(0.1,0.9)$,}
    then %given by Eq.~($\ref{eq:Mm:Vhat:3}$)
    the function $g(x,V)$ from Eq.~(\ref{def_g}) decreases over the interval
    \textcolor{black}{$(0,V_-(x))$}, increases across the range \textcolor{black}{$(V_-(x),V_+(x))$}, and subsequently decreases monotonically with $V$.
    %as illustrated in Fig.~{\ref{fig:A2}}.
    %(see, for example, the trace of $g(x,V)$, associated to the label $x=0.792$, in Fig. \ref{fig:A2}, to be discussed in detail later on).
    %,
    %and to be discussed
    %in more detail
    %} %, %inspecting , it can be realized that
    %for any state value within the set $(0.1,0.9)$
    %).

In this scenario, the optimal switching control voltage $\hat{V}(x)$ for a given state $x$ depends upon the relative positions of the levels $V_-(x)$, $V_1$, and $V_2$ along the $V$ axis.
If $V_1<V_-(x)$,
%(as may be evinced via eye inspection, this happens in Fig.~\ref{fig:A2}, to be discussed in detail later on, %inspecting , it can be realized that
%for any state value within the set $(0.3,0.4)$),
%
the following choice should be made for the optimal control voltage $\hat{V}(x)$: %for the given state $x$:

    \begin{equation}
        \hat{V}(x)=\begin{cases}
            V_2 &\textnormal{if} \qquad V_2 <  V_-(x)\;, \\
            V_-(x) &\textnormal{if} \qquad V_-(x) \leq  V_2 \leq V^*(x)\;, \\
            V_2 &\textnormal{if} \qquad V^*(x) <  V_2\;, \\
        \end{cases} \label{eq:Mm:Vhat:6}
    \end{equation}

\noindent where $V^*(x)$ is defined as the \textcolor{black}{state-dependent solution for $V$ satisfying} %voltage level such that solution to
the identity %\textcolor{black}{$g(x,V_-(x))=g(x,V^*(x))$}.
\textcolor{black}{$g(x,V_-(x))=g(x,V)$}. %see Fig.~\ref{fig:A2}.
If, on the other hand, $V_1>V_{-}(x)$ (as may be evinced via eye inspection, this happens in Fig.~\ref{fig:A2}, %to be discussed in some depth shortly, %inspecting , it can be realized that
for any value assigned to the memory state $x$ from the set $[0.1,0.9]$, then

    \begin{equation}
        \hat{V}(x)=\begin{cases}
            V_1 &\textnormal{if} \qquad \textcolor{black}{g(x,V_1) <  g(x,V_2) \;,} \\
            V_2 &\textnormal{if} \qquad \textcolor{black}{g(x,V_2) <  g(x,V_1)}. \\
        \end{cases} \label{eq:Mm:Vhat:7}
    \end{equation}

Importantly, as $x$ increases from $x_i$ to $x_f$ during a SET transition, the optimal state-dependent solution $\hat{V}(x)$ could switch in some complex form between constant levels, specifically the maximum possible voltage $V_2$, a state-dependent value $V_-(x)$, and the minimum possible voltage $V_1$,
as described via the triplet of Eqs.~(\ref{eq:Mm:Vhat:5}), (\ref{eq:Mm:Vhat:6}), and (\ref{eq:Mm:Vhat:7}).
\textcolor{black}{In particular, the most energetically-favorable control voltage $\hat{V}(x)$ is given by Eq.~(\ref{eq:Mm:Vhat:5}) if the polarity of $G_M(x)/\eta_S^2-\lambda_1$ is negative, and by
either Eq. (\ref{eq:Mm:Vhat:6}) or Eq. (\ref{eq:Mm:Vhat:7}), depending upon the relative positions of the levels $V_-(x)$, $V_1$, and $V_2$ along the $V$ axis, if the polarity of $G_M(x)/\eta_S^2-\lambda_1$ is positive.}
%becomes positive or negative, respectively.
% with reference to Eqs. (\ref{eq:Mm:Vhat:6}) and (\ref{eq:Mm:Vhat:7}), as the state $x$ visits any range of values, where $G_M(x)/\eta_S^2-\lambda_1$ keeps strictly positive,  between three different levels, due to specific variations in the relative positions of the levels $V_2$, $V_-(x)$, and $V_1$ along the $V$ axis.     %
%
\textcolor{black}{Regardless of the dependence of the optimal control voltage $\hat{V}$ upon $x$,
%which could switch between different constant levels at each state $x$, where $G_M(x)/\eta_S^2-\lambda_1$ changes polarity,
%Whether the control voltage contains a single case or multiple cases in Eq.~(\ref{eq:Mm:Vhat:6}),
the parameter $\lambda_1$ may be uniquely determined %appropriately
%determined %in a %self-consistent
%suitable way
from the state integral of the reciprocal of the % on
state evolution function $f(x,V)$, as reported in Eq.~(\ref{eq:5a}), where
%$V$ is the state-dependent function $\hat{V}(x)$
%to ensure
the effective switching time $T_c$
%, estimated by computing the state integral of the reciprocal of the % on
%state evolution function $f(x,V)$ via Eq.~(\ref{eq:5a}),
%, approximated by the first term on the right-hand side of Eq.~(\ref{eq:Mm2}),
%while disregarding second addend, in (\ref{eq:5a}),
%to be identically equal to the programming time
is known, being identically equal to $T$, as follows from Eq.~(\ref{eq:cond3}), employing the methodology outlined in the description of case 3 from the analysis aimed to minimize Joule losses in the programming of a memristor the VTEAM model was adapted to (see section \ref{sec:3a2}).}
\end{itemize}

All in all, in case 3 the proposed optimization protocol recommends the application of a sequence of consecutive pulses, whose widths and heights
%from the admissible range $[V_1,V_2]$
depend upon the time evolution of the state $x$ throughout the programming time.

\subsection{Discussion}\label{sec:A4}
Fig.~\ref{fig:A3} presents the results of applying Pontryagin’s principle to optimize the energy efficiency of the SET programming process in memristors modeled by the dynamic balance model.
The proposed optimization strategy identifies the most energetically-favorable voltage stimulus $\hat{V}(x)$ for any value assumed by the internal state $x$ as it increases from some initial value $x_i$ to some final value $x_f$ within the existence domain $[0,1]$, based on the available programming time $T$, for each of two possible SET control regimes.
As shown in the legends of plots (a) and (b) of Fig. \ref{fig:A3},  % respectively.
which of the two viable SET control regimes
should be selected depends upon the polarity of the exponential polynomial $V_1^2 e^{-\eta_SV_1} - V_2^2 e^{-\eta_SV_2}$,
%$p(V_1,V_2)$, defined as
%
%\begin{eqnarray}
%p(V_1,V_2) \triangleq V_1^2 e^{-\eta_SV_1} - V_2^2 e^{-%\eta_SV_2}, \label{poly_for_Fig_A3}
%\end{eqnarray}
%
%\noindent and
which is determined in its turn
by the choice for the lower and upper bounds in the admissible range $[V_1,V_2]$ of voltages applicable across the device.

%
%
%Like the VTEAM model, we explore different scenarios based on the allowed time interval $T$, which seems to be the most effective way to illustrate this.
%
%Specifically, we will describe two control regimes, determined by the values of $V_1$ and $V_2$.
%First of all,
Let us start off with the description of the marginal scenarios. %may occur.
In regard to the first one,
%with reference to plots (a) and (b) of Fig.~\ref{fig:A3},
irrespective of the sign of $V_1^2 e^{-\eta_SV_1} - V_2^2 e^{-\eta_SV_2}$ (refer to both plots in Fig.~\ref{fig:A3}), when the programming time $T$ is set to $T_S(V_2)$, computable via %Eq. (\ref{eq:Mm:cond1}), i.e. via
Eq. \eqref{eq:MS2} for $V_0=V_2$, and equivalent to the shortest SET programming time
%ON switching time
$T_S^{min}$, the most energetically-favorable control voltage for inducing a SET transition across the device consists of a square pulse of the highest possible height $\hat{V}$, fixed to $V_2$,
and width $T_c= T$.

In regard to the second one, when $V_1^2 e^{-\eta_SV_1} - V_2^2 e^{-\eta_SV_2}<0$ (refer to Fig.~\ref{fig:A3}(b)), and the programming time $T$ is chosen equal to $T_S(V_1)$,
a square voltage pulse of height fixed to $\hat{V}=V_1$ and width $T_c= T$
should be applied across the device to turn it on in the least power-consuming form.

Proceeding now with the analysis of the most probable situations when $T>T_S(V_2)$, and
%discussing first
considering first the scenarios illustrated in Fig.~\ref{fig:A3}(a), referring to the SET control regime, where the exponential polynomial
%$p(V_1,V_2)$ from (\ref{poly_for_Fig_A3})
$V_1^2 e^{-\eta_SV_1} - V_2^2 e^{-\eta_SV_2}$
is non-negative,
the optimal input voltage $\hat{V}$
%, under the assumption that the programming time $T$ is set to a value larger than $T_S(V_2)$,
%inequality $T>T_S(V_2)$ holds true,
consists of a rectangular pulse of height fixed to $V_2$ and width $T_c=T_S(V_2)$,
%computed via equation~(\ref{eq:Mm:cond1}), and
numerically
%equal to (smaller than)
smaller than the programming time $T$ itself.
\textcolor{black}{Here $\hat{V}$ would be forced %necessarily
%kept equal
to $0$ over the $(T-T_c)$-long remainder of the programming phase. }
%, according to the specifications set in case 2 of section \emph{A.C} %\ref{sec:A3},
%lambda_1
% turning points w,t
% T_C same for one fixed solution except for solutions depending on T
%select T_C<T where it would be possible T_C=T as well
%appendix section labels
%provided the programming time $T$ is larger than the minimum ON switching time $T_S(V_2)$.
%No solution is possible if $T<T_S(V_2)$, while for $T=T_S(V_2)$ the optimal voltage $\hat{V}$ should consist of a square voltage pulse of height $V_2$ and width $T_c=T_S(V_2)\equiv T$.
%This solution is valid for $T\geq T_S(V_2)$. If not, a solution is absent.
%
%In the case of Fig.~\ref{fig:A3}(b),
% da qui

Considering now the scenarios illustrated in Fig.~\ref{fig:A3}(b), referring to the SET control regime, where the exponential polynomial
%$p(V_1,V_2)$ from (\ref{poly_for_Fig_A3})
$V_1^2 e^{-\eta_SV_1} - V_2^2 e^{-\eta_SV_2}$ is strictly negative,
%the optimal input voltage $\hat{V}$
the recommendation for the optimal control voltage $\hat{V}$, here amenable to state dependence,
%resulting from the application of the control scheme,
depends upon the programming time. \\ %$T$. \\
%Let us assume $T$ to be larger than
%satisfies the inequality which must mandatorily be larger than
%$T_S(V_2)$.
%is more involved.
% da qui
%The main idea is that, in this case,
If $T\in[T_S(V_2),T_S(V_1)]$,
%it is chosen from the range $(T_S(V_2),T_S(V_1))$,
the optimal voltage $\hat{V}(x)$ may in general switch
between different levels. In particular, if the polarity of $G_M(x)/\eta_S^2-\lambda_1$ is negative, $\hat{V}(x)$ would be set as dictated by Eq.~(\ref{eq:Mm:Vhat:5}), whereas, otherwise, it would be set according to
%as dictated by
either Eq. (\ref{eq:Mm:Vhat:6}) or Eq. (\ref{eq:Mm:Vhat:7}), depending upon the relative positions of the levels $V_-(x)$, $V_1$, and $V_2$ along the $V$ axis.
On the other hand, if $T>T_S(V_1)$,
%is set to some value beyond $T_S(V_1)$,
%the minimum time to switch the device on as the lowest possible voltage $V_1$ is let fall across it,
the optimal switching control protocol recommends to apply a voltage stimulus
$\hat{V}(x)$ identically equal to $V_1$ for a time interval $T_c$ equal to $T_S(V_1)$, computable via Eq.~(\ref{eq:MS2})
% Fig.~(\ref{eq:Mm:cond1})
for $V_0=V_1$, \textcolor{black}{while no input should further stimulate the device in the $(T-T_c)$-long remainder of the programming phase. } %\\

With reference to the control regime, illustrated in Fig. \ref{fig:A3}(b), the necessity to endow the exponential polynomial
$V_1^2 e^{-\eta_SV_1} - V_2^2 e^{-\eta_SV_2}$ with a negative polarity enforces a constraint on the choice of the lower bound in the range of voltages, which may be let fall across the memristor. In fact $V_1$
must necessarily be
%needs to be
smaller than the abscissa $2/\eta_S$ of the maximum of the function \textcolor{black}{$V^2 e^{-\eta_SV}$}.
For example, for $\eta_S=5$~V$^{-1}$, the inequality $V_1<0.4$~V
must mandatorily hold true.

\textcolor{black}{
As the programming time $T$ is progressively increased from $T_S^{min}=T_S(V_2)$, the SET switching energy $Q_S$ keeps equal to its smallest possible value in the control regime from panel (a), whereas it decreases monotonically from its largest possible value to its smallest possible value in the control regime from panel (b), similarly as was the case in Fig. \ref{fig:7} for the VTEAM model.
}

\begin{figure}[tb]
\centering
\includegraphics[scale=0.31]{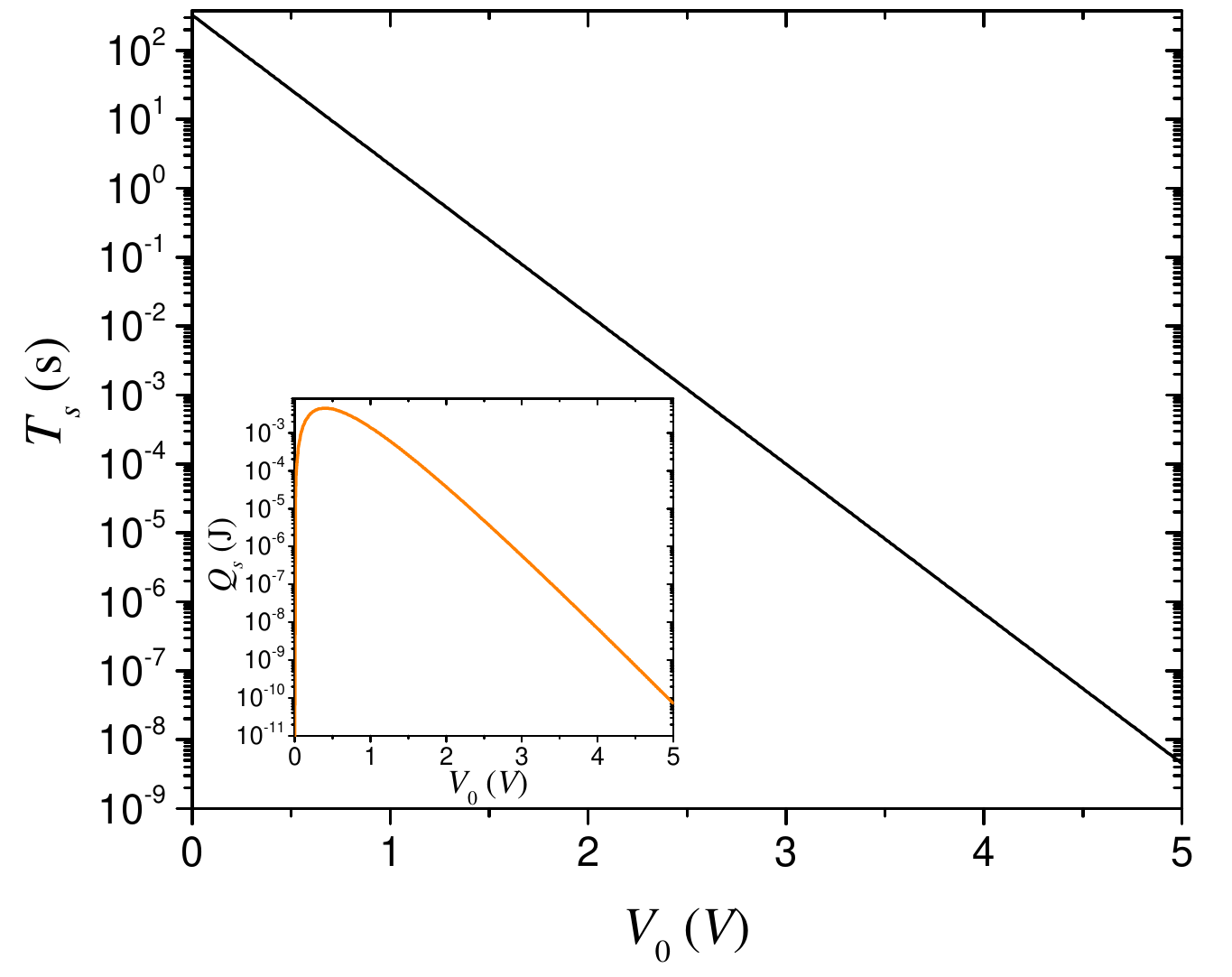}
\caption{Time $T_S$ and energy $Q_S$
%Joule losses
(inset), necessary to induce an increase in the device internal state $x$ from $x_i=0.1$ to $x_f=0.9$, as a function of the pulse height $V_0$, as predicted via the approximate form of the dynamic balance model, and
%Eqs.~(\ref{eq:MS2}) and (\ref{eq:MS3}), respectively,
%while neglecting the second RESET term in the sum on the right-hand side of Eq.~(\ref{eq:Mm2}),
assuming the very same values for $\eta_S$, $\tau_{0,S}$, $G_{min}$, and $G_{max}$ as reported in the caption of
%in the simulation
Fig. \ref{fig:A5}.} % is associated to.
%for the very same following
%parameter set: $\eta_S=5$~V$^{-1}$, $\tau_{0,S}=148$~s, $G_{min}=1$~{\textmu}S, and $G_{max}=1$~mS.
%
%(refer to Fig. \ref{fig:A5} also).
%}
%This figure was obtained using Eqs.~(\ref{eq:MS2}) and (\ref{eq:MS3}) with the following set of parameters: $\eta_S=5$~V$^{-1}$, $\tau_{0,S}=148$~s, $x_i=0.1$, $x_f=0.9$, $G_{min}=1$~{\textmu}S, and $G_{max}=1$~mS.}
\label{fig:A4}
\end{figure}

\begin{figure}[tb]
\centering
\includegraphics[scale=0.31]{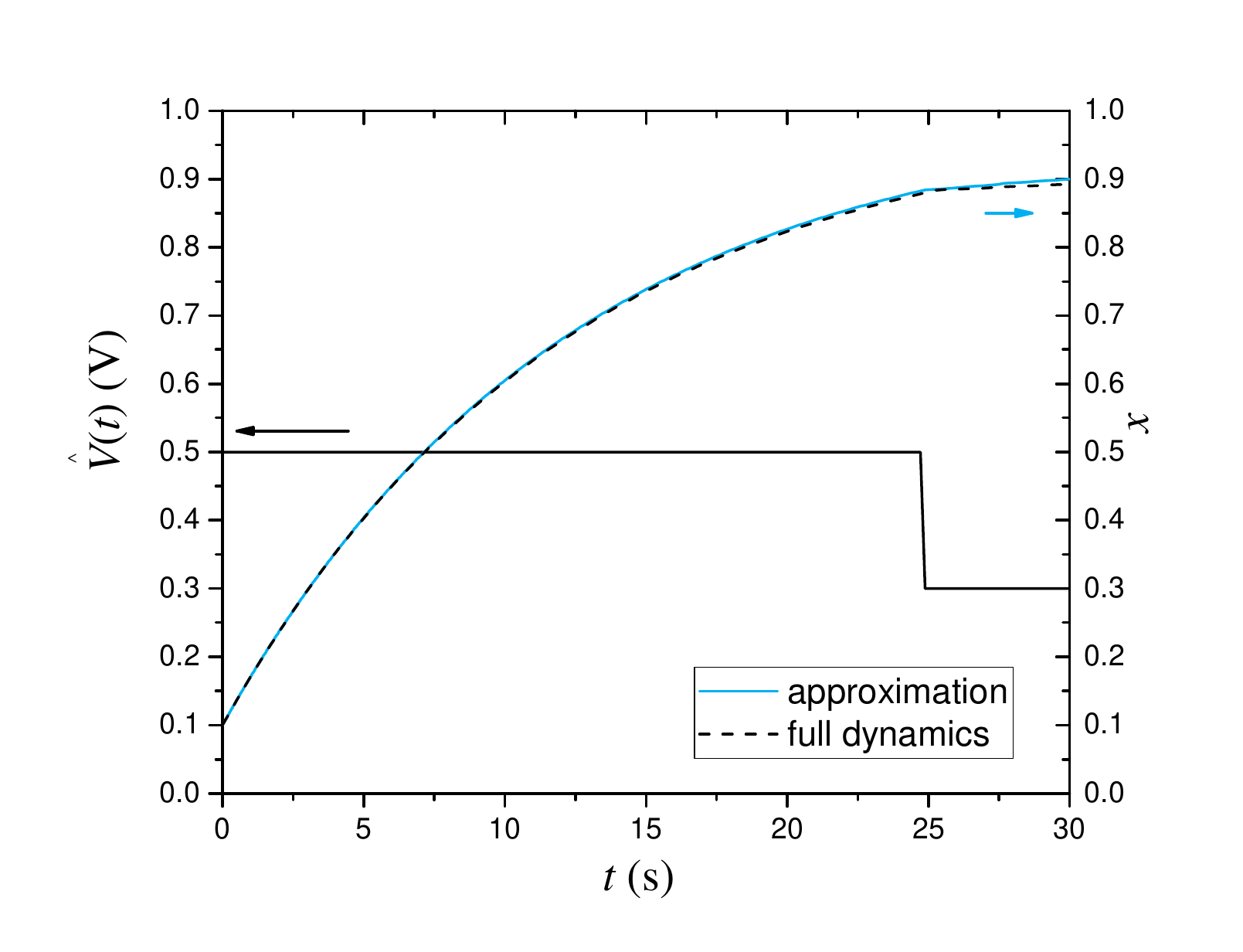}
\caption{Application of the proposed optimal SET switching control protocol to increase the device state $x$ from $x_i=0.1$ to $x_f=0.9$ in a scenario, where %the exponential polynomial
\textcolor{black}{$V_1^2 e^{-\eta_S V_1}-V_2^2 e^{-\eta_S V_2}$} is enforced to be negative % strictly
%negative, as follows
upon setting $\eta_S$ to $5$~V$^{-1}$, $V_1$ to $0.3$~V, and $V_2$ to $0.5$~V, which calls for the need to consider case 3 from section \emph{A.C}, $T$ is set to $30$~s, thus
%under the additional hypothesis that the %programming time
%$T$ %72.34~\mu\text{s}$,
%$T = 72.34~\mu\text{s}$, $\lambda_1 = 12~\mu\text{S}\,\text{V}^2$, $\lambda_0 = 1$.
lying inside the range $(T_S(V_2),T_S(V_1))=(26.69$~s$,72.56$~s$)$, %
%T_S(V_2=0.7 V)=9.82 s,
%T_S(V_1=0.1 V)=197.24 s,
%T=T_c=72.34 s
%being
%fixed %set beforehand
%to $30$~s,
\textcolor{black}{$\lambda_1$ is found to be equal to $2.7515$~{\textmu}SV$^2$ via Eq. \eqref{eq:5a},}
%With
$G_{min}$ and $G_{max}$ were respectively set to $1$~{\textmu}S and to $1$~mS, whereas $\tau_{0,S}$ was taken equal to $148$~s. %the optimal pulse sequence for $\hat{V}$,
%based on case 3 from Section \emph{A.C},
%
%at all times.
%Other model parameters:
%(the formula (\ref{eq:5a}) for the \textcolor{black}{effective} switching time $T_c$,
%coinciding with the programming time $T$,
%was used to find a suitable value for $\lambda_1$, specifically
%is self-consistently set to
%$2.7515\,\mu\text{S}\cdot\text{V}^2$).
%$12~\mu\text{S}\,\text{V}^2$).
%%UPDATE 18 August 2025
%$V_1 = 0.3 \, \text{V}$ , $V_2 = 0.5 \, \text{V}$,  $T = 30 \, \text{s}$ ($\lambda_1 = 2.75$),  provides a solution to the problem since $T=30 \, \text{s} \ll \tau_0 = 148 \, \text{s}$.
%%
Black (light blue) solid trace: Optimal control voltage $\hat{V}$ (memristor %internal
state $x$) over time, as descending from our energy-saving programming procedure. % proposed
%optimal control protocol.
%
%
%The current data point of either of the two traces affects the next data point of the other trace.
%In fact, at any given time instant, the optimal input voltage is determined according to the triplet of Eqs.~
%(\ref{eq:Mm:Vhat:5}), (\ref{eq:Mm:Vhat:6}), and (\ref{eq:Mm:Vhat:7}) on the basis of the value assumed by the state at the previous time point.
%Similarly, the current state value is determined parametrically via Eq.~(\ref{eq:5a}) on the basis of the previous input voltage value.
%Moreover, in these calculations the RESET term is omitted from the formula of the state evolution function $f(x,V)$ appearing on the right-hand side of Eq.~(\ref{eq:Mm2}).
%In particular, as $x$ increases from $x_i=0.1$ to $x_f=0.9$, $g(x,V)$ maintains a non-monotonic shape (see Fig. \ref{fig:A2}), and $V_-(x)$ keeps smaller than $V_1$. As a result, for the parameter setting under test ($\eta_S=5$~V$^{-1}$,
%$\tau_{0,S}=148$~s,
%$G_{min}=1$~{\textmu}S, $G_{max}=1$~mS), $\hat{V}$ is effectively governed only by Eq.~(\ref{eq:Mm:Vhat:7}), switching from $V_2=0.5$V to $V_1=0.3$V at $t = 24.7\,$\text{s}.
Black dashed trace: Memristor state $x$ over time, as resulting from the numerical integration of the original ODE (\ref{eq:Mm2}) for $\tau_{0,R}=\tau_{0,S}=148$~s and $\eta_R=-\eta_S=-5$~V.
% DAE set composed of equations
% (\ref{eq:Mm2}) and (\ref{eq:Mm1})
from $x_i=x(t_i)=0.1$, with $t_i=0$, and under the very same input voltage $\hat{V}$ --  see the black solid trace once again -- recommended by the optimization process.
%, and indicated, as mentioned earlier, through the black solid trace.
%The parameter setting for this numerical simulation follows:
%$\eta_S=-\eta_R=5$~V$^{-1}$,
%$\tau_{0,S}=\tau_{0,R}=148$~s,
%$G_{min}=1$~{\textmu}S, $G_{max}=1$~mS.
}
%
%As explained in the text, the time evolution of $\hat{V}$ is affected by the dynamics of the internal state, as governed by the triplet of equations (\ref{eq:Mm:Vhat:5}), (\ref{eq:Mm:Vhat:6}), and (\ref{eq:Mm:Vhat:7}).
%
%The values of all the model parameters, which have not been mentioned in the caption, are the same as in Fig. \ref{fig:A4}.}
%
%Dynamic balance model: An example of optimal control corresponding to the central regime in Fig.~\ref{fig:A3}(b).
%In this example, the control voltage $\hat{V}$ changes from $V_2=0.7$~V to
%$V_1=0.1$~V at $t\approx 6.54$~s.
%All other  parameters are the same as in Fig.~\ref{fig:A4}.}
\label{fig:A5} \vspace{-0.3cm}
\end{figure}

As a
proof of evidence for the accuracy of the results, acquired through the application of the proposed SET switching energy minimization
technique to the dynamic balance model, we conclude this section with the analysis of a scenario, which falls in case 3 from section \emph{A.C}, since the exponential polynomial $V_1^2 e^{-\eta_S V_1}-V_2^2 e^{-\eta_S V_2}$ was enforced to be negative upon choosing $\eta_S=5$~V$^{-1}$, $V_1=0.3$~V, and $V_2=0.5$~V.
Looking at Fig. \ref{fig:A4}, showing the dependence of the SET switching time $T_S$ upon the pulse height $V_0$ according to the formula (\ref{eq:MS2}), with
$x_i$ and $x_f$ respectively set to $0.1$ and $0.9$,
%while neglecting the second additive RESET term on the right-hand side of Eq.~(\ref{eq:Mm2}),
%
$T_S(V_2)$ ($T_S(V_1)$) is found to be equal to $26.69$~s ($72.56$~s).
%(for the sake of completeness, the inset of Fig. \ref{fig:A4} shows the locus of the SET switching energy $Q_S$ upon the pulse height $V_0$ according to the formula (\ref{eq:MS3})).
%
Taking the programming time $T$ equal to $30$~s, which lies inside the range $[T_S(V_2),T_S(V_1)]$, %=(26.69$s$,72.56$s$)$.
setting $G_{min}$ and $G_{max}$ to $1$~\textmu S and to $1$~mS, respectively, and choosing a value of $148$~s for $\tau_{0,S}$, \textcolor{black}{the formula %Eq.
\eqref{eq:5a} for $T_c$, here equal to $T$,
was employed to determine a unique value for $\lambda_1$, specifically $2.7515$~\textmu SV$^2$.}
%, while, as announced at the beginning of case 3 from section \emph{A.C}, $\lambda_0$ was set to $1$.
As a result, the state-dependent function $G_M(x)/\eta_S^2-\lambda_1$, where $G_M(x)$ is inferable from Eq. (\ref{eq:Mm1}), was found to keep positive, irrespective of the value of the memristor state $x$ across the range $[x_i,x_f]=[0.1,0.9]$.
%Moreover,
It follows that,
%As a result,
as $x$ increases from $x_i=0.1$ to $x_f=0.9$, the locus of $g(x,V)$ versus $V$ maintains a non-monotonic shape (refer to Fig. \ref{fig:A2}), %while, concurrently,
admitting a local minimum, which keeps consistently smaller than $V_1$, when $V$ assumes the value $V_-(x)$, whose formula is reported in Eq. (\ref{eq:Mm:Vhat:4}). % \\
%$V_1-V_-(x)>0$
%
As a result, the choice for the optimal pulse sequence-based
voltage stimulus $\hat{V}(x)$ for the least power-consuming SET operation fell for Eq. \eqref{eq:Mm:Vhat:7}. As shown in Fig. \ref{fig:A5}, while $x$ (solid light-blue trace) increased from $0.1$ to $0.9$, the sign of the inequality $g(x,V_1)-g(x,V_2)$ was found to switch from positive to negative at $t = 24.7$~s, when
%In the simulation Fig. \ref{fig:A5} refers to, as governed by Eq.~(\ref{eq:Mm:Vhat:7}),
the optimal voltage $\hat{V}$ (solid black trace) %was found to undergo a single
transitioned %over the programming phase, specifically
from its initial value $V_2=0.5$~V to its final value $V_1=0.3$~V.
%at $t = 24.7\,$\text{s}, when $g(x,V_2)-g(x,V_1)$ turns from negative to positive.
%The control voltage (state sample) at any point in time depends upon the state (control voltage) sample at the previous point in time according to Eq.~(\ref{eq:Mm:Vhat:7}) (\ref{eq:5}).
%In order for the degradation in accuracy, descending from the omission of the second RESET addend on the right-hand side of Eq. \eqref{eq:Mm2}, to keep negligible, $\tau_{0,R}$ was set equal to $\tau_{0,S}$, which is way larger than $T$.

Let us set  $\tau_{0,R}$  to the same value as  $\tau_{0,S}=148$~s, which is much longer than the programming time $T=30$~s. Under this condition, during the SET transition, the second RESET term on the right-hand side of Eq.~\eqref{eq:Mm2} is  small. As a consequence, neglecting this term has a negligible impact on the dynamics of the system. This is confirmed in Fig.~\ref{fig:A5}, where the dashed black trace represents the solution of the ODE~\eqref{eq:Mm2} in its original form, obtained from the same initial condition and under the same voltage stimulus, while the solid light blue trace corresponds to the solution derived from the approximate model using the optimization technique. The two solutions are virtually indistinguishable.

%As anticipated earlier, in order to justify the omission of the second RESET addend on the right-hand side of Eq.~\eqref{eq:Mm2} during the analysis of the SET transition, i.e. under finite positive-valued voltage stimuli, $\tau_{0,R}$ was set to the same value as $\tau_{0,S}$, i.e. $148$~s, which is much larger than the programming time $T$, here chosen equal to $30$~s.
%The dashed black trace in Fig. \ref{fig:A5} shows the solution to the ODE \eqref{eq:Mm2} in its original form from the same initial condition and under the same voltage stimulus, which resulted in the solid light blue trace. Clearly the solution, obtained by applying the optimization technique to the approximate model, deviates negligibly %imperceptibly
%from the exact one.

Compared to the case where a single SET pulse with duration $T_S=30$~s and amplitude $V_0=0.477$~V, determined from Eq. (\ref{eq:MS2}), is used to switch the device on by raising its state from $0.1$ to $0.9$, the Joule losses in the optimal-control-based switching scheme are reduced by a factor of $1.023$. \textcolor{black}{While the improvement is small, this value is not representative of our approach overall.}

A similar classification could be conceived for the
voltage stimuli, resulting in the least power-consuming RESET %switching
transitions across the device, the dynamic balance model is assumed to be adapted to, on the basis of the choice for the programming time $T$, and depending upon the values
of the least and most negative voltages, namely $V_1'$ and $V_2'$ respectively, applicable across the memristor.
%Importantly, the approximation, descending from the omission of the RESET term in the state equation \eqref{eq:Mm2} during the theoretical procedure proposed in this paper for identifying a suitable control voltage allowing to minimize the energy costs associated with the device SET transition, introduces a negligible accuracy degradation, provided the programming time $T$ is considerably smaller than the RESET time constant $\tau_{0,R}$.
%Assume $\tau_{0,S}$ and $\tau_{0,R}$ to be equal one to the other and as large as $148$~s, and let $\eta_R=-\eta_S=-5$~V$^{-1}$, while taking
%$G_{min}$ and $G_{max}$ equal to $1$~{\textmu}S and to $1$~mS, %respectively.
%Setting $V_1$ ($V_2$) to $0.3$~V ($0.5$~V),
%
%Assigning a value of $30$~s to the programming time $T$,
%are long enough to satisfy the hypothesis $\tau_{0,S},\tau_{0,R}<<T$
%
%forming the
%has a direct consequence on the choice of the SET and RESET time constants, which, according to
%introduced in section \emph{A.A}.}
%should satisfy
%the inequality
%i.e. $\tau_{0,S},\tau_{0,R}<<T_S(V_1)$ is satisfied.}.
%%
%Setting furthermore $V_2$ to $0.7$~V, which implies $T_S(V_2)=9.82$~s, as may be read from Fig. \ref{fig:A3}, and assigning a value of $72.34$~s to the programming time $T$, the situation under test may be classified as a scenario from case 3 of section \emph{A.C}.
%
%
%
%}
%
%missing points after lunch: captions last two figures, replies to reviewers, point from Yuriy, discussion with Valery on \lambda_1
%%

\end{document}